\documentclass[11pt,a4paper]{article}
\usepackage{color}
\usepackage{amsmath,amsthm,amssymb,color,array,multirow}
\usepackage{epsfig,amsfonts,latexsym, hyperref,  mathrsfs}
\usepackage{natbib}
\usepackage{tikz}
\usetikzlibrary{shadows,trees}
\usepackage{amscd}
\usepackage{fancyhdr,latexsym,amsmath,amsfonts,amssymb,amsbsy,amsthm,url}%
\usepackage{natbib}
\usepackage[hscale=0.7,centering]{geometry}
\usepackage{graphicx,epsfig, xy}
\usepackage{hyperref}
\usepackage{lscape,fancybox}
\usepackage{color}
\usepackage{caption}
\usepackage{subcaption}

\newcommand{\blmma}{\begin{lemma}}
\newcommand{\elmma}{\end{lemma}}

\newtheorem{lemma}{Lemma}

\numberwithin{equation}{section}

\title{\textbf{Spatio-temporal patterns of diurnal temperature:  a random matrix approach
\\  
I-case of India\\
}   
}

\author{Arup Bose}
\date{}

\begin{document}
	\def\shortauthors{ABC}
	\author{
		\hspace{0.01\textwidth}
		\parbox[t]{0.50\textwidth}{{Madhuchhanda Bhattacharjee}
				\\ Department of Mathematics\\{University of Manchester\\
				chhanda.bhatta@gmail.com\\
					\\ }}
		\hspace{0.01\textwidth}
		\parbox[t]{0.5\textwidth}{{Arup Bose}\thanks{Research  supported by J.C.~Bose National Fellowship, JBR/2023/000023 from Science and Engineering Board, Govt.~of India.}
			\\ Statistics and Mathematics Unit\\
			Indian Statistical Institute\\
				bosearu@gmail.com
				}
		}
		\maketitle

\begin{abstract} We consider the spatio-temporal gridded daily diurnal temperature range (DTR) data across India during the 72-year period 1951--2022. We augment this data with information on the El Ni\~no-Southern Oscillation (ENSO) and on the climatic regions (Stamp’s and Koeppen’s classification) and four seasons of India.

We use various matrix theory approaches to trim out strong but routine signals, random matrix theory to remove noise, and novel empirical generalised singular-value distributions to establish retention of essential signals in the trimmed data. We make use of the spatial Bergsma ($S_B$) statistics to measure spatial association and identify temporal change points in the spatial-association.

In particular, our investigation captures a yet unknown change-point over the 72 years under study with drastic changes in spatial-association of DTR in India. It also brings out changes in spatial association with regard to ENSO. 

We conclude that while studying/modelling Indian DTR data, due consideration should be granted to the strong spatial association that is being persistently exhibited over decades, and provision should be kept for potential change points in the temporal behaviour, which in turn can bring moderate to dramatic changes in the spatial association pattern. 

Some of our analysis also reaffirms the conclusions made by other authors, regarding spatial and temporal behavior of DTR, adding our own insights. We consider the data from the yearly, seasonal and climatic zones points of view, and discover several new and interesting statistical structures which should be of interest, especially to climatologists and statisticians. Our methods are not country specific and could be used profitably for DTR data from other geographical areas. 
\end{abstract}

\vskip5pt
	

\section{Introduction} 
Let $T_{max}$ and $T_{min}$ denote the maximum and minimum near surface air temperatures within a day (a period of 24 hours). Then the \textit{Diurnal Temperature Range} (DTR) for that day is defined as $T_{max}-T_{min}$. Recorded DTR data over two centuries across the globe has become an increasingly important spatio-temporal historical record, especially due to global warming. It is acknowledged that changes in many environmental indices are causally linked to changes in local and global DTR values. Variations in the DTR has significant impact on life and agriculture. It is important to understand the space-time behaviour of DTR across locations over time. 

Time series of \textit{yearly average} DTR from different regions across the world have been investigated by several authors from various perspectives, \cite{zhong2023}, \cite{sharma2021}, \cite{stjern2020}, \cite{mall2021}, \cite{roy2019}, to mention a few. Some of these articles study DTR at global level (for example  \cite{zhong2023} and \cite{stjern2020}), while others make specific contributions on knowledge of DTR for the Indian subcontinent (for instance \cite{sharma2021} \cite{stjern2020}, \cite{mall2021}, \cite{roy2019}, \cite{vinnarasi2017}, \cite{jhajharia2011} and \cite{kothawale2010}).


All of these studies primarily focus on the \textit{temporal pattern} of DTR. A few do recognise the existence of spatial variation (that is, variations across space) and present some results at individual spatial units, like grids. Terminologies such as ``spatial distributions" and ``spatial variations'' that appear in some of these articles, always are synonymous with presentation of the results in the form of a map. It is typically left to the reader to gain any insight into possible ``spatial" variation. Since the DTR is measured at specific times and locations, terms like ``spatial distribution" might be a bit misleading, since that would normally refer to some randomness in the space variable. The only randomness here is in the magnitude of the DTR.

The randomness in DTR is expected to have strong temporal as well as spatial components. However the  literature appears to have focused extensively on studying the temporal behaviour of DTR, without attempting to capture the inherent spatial dependence.
Any such attempt should involve the spatial information of the grids. 

 Admittedly the temporal component has a strong presence which masks the other contributors like spatial information. \cite{wu2009} developed \textit{Multidimensional Ensemble Empirical Mode Decomposition} (MEEMD) and this was applied by \cite{ji2014} to climate data for eliminating the oscillatory component of a time series and potentially reveal the slow varying components. \cite{vinnarasi2017} used this technique to Indian DTR data. However, such techniques do not account for interdependence of the DTR series arising from various geographic locations.

We consider the DTR data for India, which is available since 1951. We try to extract the interdependence of the spatial units with regard to DTR. Since the common temporal factors make it difficult to isolate such salient patterns, we employ novel linear (random) matrix theory techniques to de-trend the data temporally, while preserving essential spatial signals. This technique allows us to study the time series jointly rather than detrend them individually. We employ a second novel technique to quantify the extent of spatial (statistical) dependence by a univariate statistics. This eliminates the need of having to rely on subjective judgements based on maps to arrive at an understanding of the degree of spatial association/dependence. Our method allows implementation at various temporal units and windows, which enables us to study the temporal change in such spatial dependence. This lead us to uncover yet unknown striking changes in spatial association pattern of DTR across India.

We emphasise that though we focus on India, our methods are not region specific and are applicable mutatis mutandis to other regions of the world. The analysis pipeline proposed here can be implemented using standard software like {\tt R} for any similar cohort of time series that arise from a collection of geographical units.

Section \ref{sec:materials_methods} on materials and methods is divided into two sub-sections. In Section \ref{sec:materials}, after describing the data, which is spatio-temporal in nature, we discuss the issues related to organising it appropriately in a matrix form. 
Then in Section \ref{sec:methods} we describe in brief the ideas that we have used from matrix and random matrix theory. These include the regular and generalised singular-value decompositions, empirical spectral distribution, Mar\v{c}enko-Pastur law. We also describe Spatial-Bergsma statistic which we will use as a statistical measure of spatial association. 

In Section \ref{sec:preliminary_exploration} we present the result of our preliminary explorations using the above ideas. In Section \ref{sec:analysis} we dig into a very detailed analysis, and describe the results of our analysis. In Section \ref{sec:conclusions} we summarise our findings briefly. 

In Section \ref{sec:suppl_plot} we show some additional analysis of the DTR, especially under different stratifications using climatic zones and seasons. Since there are several figures in this article, in Section \ref{sec:notes_on_figures} we provide at one place the descriptions of all the figures.

\section{Materials and methods}\label{sec:materials_methods}

\subsection{Materials: the data}\label{sec:materials} 
\begin{figure}[!ht]
     \centering
     \includegraphics[scale=.6]{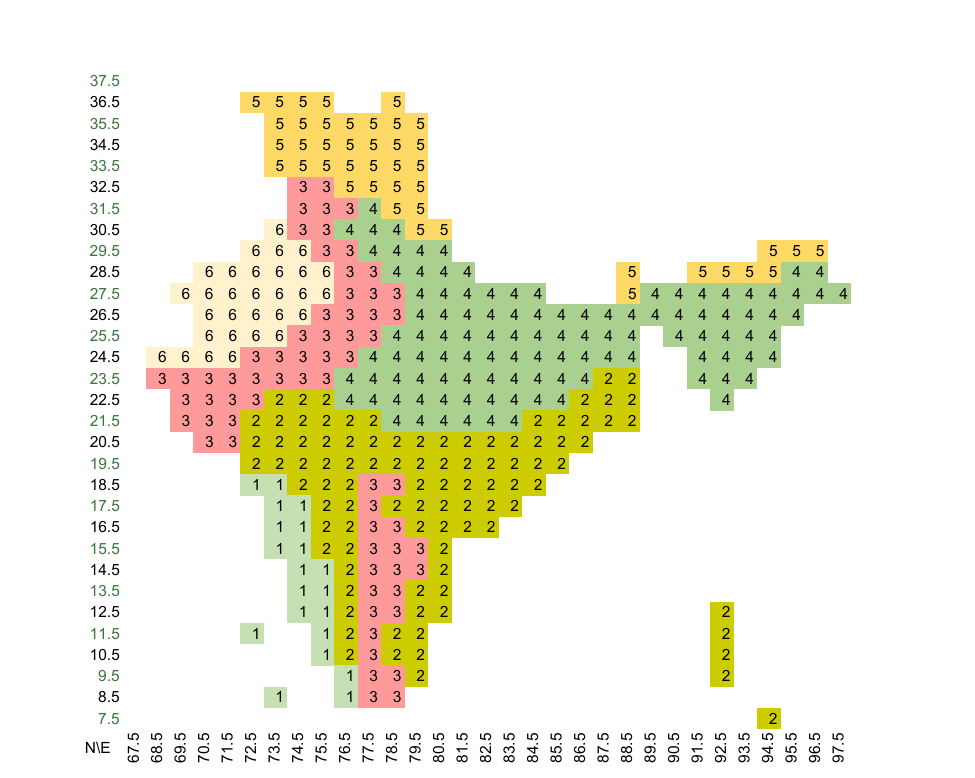}\ \ \caption{Climate grid of India: 1-Tropical monsoon; 2-Tropical savannah, wet and dry; 3-Arid, steppe, hot; 4-Humid subtropical; 5-Montane climate; 6-Hot deserts, Arid.}
    \label{fig:India_climate_map.pdf}
\end{figure}

In Figure \ref{fig:India_climate_map.pdf}, the spatial domain of India is split into a grid of 362 locations, spanning six climatic regions,  between $7.5^{\circ}$N--$37.5^{\circ}$N and $67.5^{\circ}$E--$97.5^{\circ}$E, with a spatial resolution of $1^{\circ}\times 1^{\circ}$. The daily DTR data at each location for the 72-year period 1951--2022 is sourced from Climate Research Services, India Meteorological Department, Govt. of India  (\textit{https://www.imdpune.gov.in/lrfindex.php}). Thus, we have 362 daily time series spanning 72 years, with some missing values. Complete data is available only for 280 of the 362 locations. This data could be grouped according to months/seasons/years in temporal dimension and/or according to the six climatic regions in the spatial dimension. Such grouping will of course depend on the aspects that we wish to probe. 

\subsubsection{Data matrix} The  time series of the daily DTR values for the 280 (or 362) locations for the 72 years yields a 280 (or 362) dimensional vector time series with 26298 time points. We visualise this as a $26298\times 280 \ ({\rm or} \ 362)$ large \textit{data matrix} $X$ say. For a specific year, the row dimension for the data sub-matrix is 365 or 366, and the rows are arranged in the naturally increasing order of time. For now, the 280 (or 362) locations (columns) are arranged in some fixed order (e.g. latitude x longitude), and we shall have more to say on this below. The entries of $X$ are subject to location dependent temporal patterns (e.g. seasonal variation). There is also spatial dependence across the locations, which is expected to be non-stationary across different pairs of locations. 

\subsubsection{Ordering the locations} Naively, the collection of 280 or 362 time series can be simply put side-by-side to form a data matrix for analysis. However each series arising out of a grid also has two key additional pieces of information, namely their latitudes and longitudes. Therefore, to investigate the temporal and/or spatial behaviour of the DTR across times and regions, we need to decide how this two-dimensional location information should be incorporated so as to arrive at an appropriate ordering amongst these time series. The consequent visual message might (and does) depend on this arrangement.

There are several choices for ordering the locations, dictated by convenience, geographical proximity, or division of India into the six climatic zones. One may begin with the lowest latitude, and move according to the increasing order of longitude of the locations, then come back to the next latitude and repeat the process. One could even switch the roles of the longitudes and latitudes. In any case, with this approach, two regions in the same climatic zone might end up being distant in the ordering. 

On the other hand, we could use a classification of the locations into climatic zones. The \cite{koppen} classification of climatic zones of the world was later modified by several scientists (see \cite{mcknight}). This yields a classification of India into \textit{six} major climatic zones as given in Figure \ref{fig:India_climate_map.pdf}. We could group the locations according to the climatic zones they belong to. This would imply that locations that are geographically close, are not necessarily close with respect to the ordering. One may use the latitude-longitude ordering within each zone to alleviate this problem somewhat. 

We tried several different orderings, and eventually settled on a \textit{spiral ordering} that is a type of Hilbert space-filling curve, which are sometimes used in image processing to rearrange image pixels in order to enhance pixel locality. Mathematically this can be thought of as a bijection from $\mathbb{N}^2$ to $\mathbb{N}$, where $\mathbb{N}=\{1, 2, \ldots\}$. In our context this addresses both the zonal and spatial proximity of the grids reasonably well.

Visualising the locations according to the longitudes and latitudes, we start with the lowest latitude and longitude position, say $(1,1)$ position, then move to the $(1,2)$ position, and then move down diagonally to the $(2,1)$ position, then move down to the $(3, 1)$ position, then successively to $(2,2)$, $(1, 3)$, and so on. In other words, we move anti-diagonally and wrap around. Likewise, instead of first moving to $(1,2)$, one could move first to  $(2,1)$  etc. We refer to this ordering as the \textit{spiral ordering}. This ordering provides a significantly improved visual depiction and will get reflected in our figures and plots, thereby conveying valuable insight into the data and its behaviour from the spatial point of view. \textit{We shall call this $26298\times 280$ reorganised data matrix} $D$.

\subsection{Methods}\label{sec:methods} We have used ideas from linear matrix theory, random matrix theory (RMT), and statistics to analyse the data. These include the regular and generalised singular-value decompositions (SVD and GSVD), empirical spectral distribution (ESD), Mar\v{c}enko-Pastur (MP) law, Bergsma's correlation $\rho$, and Spatial-Bergsma ($S_B$) which is a statistical measure of spatial association based on Bergsma's correlation. The following discussion is general, and will remain valid irrespective of any specific ordering of the locations. 

\subsubsection{Correlation matrices} All correlation matrices use daily data at the grid level and the correlations are calculated between grids. 
Depending on the situation, these matrices are calculated for the entire period (or for subsets of days, such as years, months within a year, etc.), and for the entire grid (or for subsets of grids such as climatic regions, etc.). 

The correlations have been computed based on three different data matrices: $D=((d_{dg})), \ T=((t_{dg}))$ and $S=((S_{dg}))$ $d =1, \ldots, 26298, \, \, \hbox{and} \, \, g = 1, \ldots, 280$, which represent respectively, the original DTR data, the residuals from time series decomposition of the original data, and the trimmed DTR data (trimming is explained later).

Two different correlations have been used. One is the \textit{product moment (Pearson's) correlation}, and the other is the \textit{Bergsma's correlation}. The latter has been used in the context of spatial association measures. Specifically, correlation matrices that have been computed are listed below: 
\vskip3pt

\noindent (1) Two Pearson correlation matrices $R^{D}$ and $R^{S}$, based on $D$ and $S$ respectively.
\vskip3pt

\noindent (2) Two sets of Pearson correlation matrices $\{R^{D}_{i}\}$ and $\{R^{S}_{i}\}$ based respectively on the $i$-th year sub-matrices of $D$ and $S$.
\vskip3pt

\noindent (3) Two sets of Bergsma's correlation matrices $\{B^{D}_{i}\}$ and $\{B^{S}_{i}\}$ based respectively on the $i$-th year sub-matrices of $D$ and $S$.  
\vskip3pt

\noindent (4) Bergsma's correlation matrices, $\{B^{S}_{i,k}\}$, based on $S$ for each year $i$, and only for grids from each climatic region $k$. 
\vskip3pt

\noindent (5) Bergsma's correlation matrix $B^{S}_{mi}$ based on $S$ for the $m$-th month of the $i$-th year and all grids. 

\subsubsection{Empirical spectral distribution; Mar\v{c}enko-Pastur law} The sample correlation matrix is a \textit{real symmetric random matrix}. The study of the eigenvalues of a random matrix occupies a central position in random matrix theory (RMT). A basic notion is the following: for any real symmetric $n \times n$ random matrix $R_n$, let $\lambda_1, \ldots, \lambda_{n}$ be its  eigenvalues (which are all real). Then the \textit{empirical spectral distribution} (ESD) of $R_n$ is defined by
\begin{equation}
F_{R_n }(x) \ =\ \frac{\#\{i \le n:
\lambda_i\le x\}}{n}.
\end{equation}
This corresponds to the probability distribution which puts mass $n^{-1}$ at each of the eigenvalues. Roughly speaking this is the \textit{histogram} of the eigenvalues of $R_n$. We shall use the ESD for comparing sample correlation matrices obtained using various contextual subsets of our data.

The \textit{Mar\v{c}enko-Pastur {\rm (MP)} law} provides the asymptotic distribution of the ESD of the sample covariance matrix when dimensions grow and the observations are independent and identically distributed (iid, pre noise). In particular, suppose $X$ is an $n\times p$ matrix with iid observations that have mean $0$, variance $1$ and finite fourth moment. Suppose $n \to \infty$ and $n^{-1}p\to y$, $0< y < \infty$. Then the ESD of $n^{-1}X^{\top} X$ (the symbol $\top$ denotes the transpose of a matrix) converges almost surely to what is called the MP law. For details, please refer to \cite{bose2021}. This law has a point mass $1-y^{-1}$ whenever $ y > 1$. Elsewhere (irrespective of the magnitude of $y$) it has the density
$$f(x)=\dfrac{1}{2\pi}\sqrt{(y_{+}-x)(x-y_{-})}, \ \ y_{\pm }=(1\pm\sqrt{y})^2, \ \ y_{-} \leq x \leq y_{+}.$$ 
Note that this limit law does not depend on the underlying distribution of the variables. Indeed, it is also known that with a high probability, there are no stray eigenvalues to the left of $y_{-}$ or and right of $y_{+}$. 

This can be used to distinguish signal from noise, by identifying the eigenvalues that are near the edges of the MP law. 
For example, if $p=280$ and $n=365$, then $y=0.767123$, 
and $y_{+}=(1+\sqrt{y})^2=3.519,\ y_{-}=(1-\sqrt{y})^2=0.0154$. 
For a random correlation matrix, the eigenvalues that lie within this range approximately may be attributed to noise, and the rest may be attributed to signal. 

Using the MP law, the original matrix can be approximated by using the significant eigenvalues and corresponding eigenvectors, which would be the de-noised version of the original matrix.

\subsubsection{Singular value decomposition (SVD)} Let $X$ be an $n \times p$  matrix with real entries. Then one can choose two real matrices $U$ and $V$ of orders $n\times n$ and $p\times p$ respectively, and   an $m \times n$ \textit{rectangular diagonal matrix} $\Sigma=((\sigma_{ij}))$ (so that $\sigma_{ij}=0$ if $i\neq j$) with real entries,  such that, the columns $\{u_i\}$ and $\{v_j\}$ of $U$ and $V$ are the (orthonormal) eigenvectors of $XX^\top$ and $X^\top X$, and 
$$UU^{\top}=I_{n\times n}, \ \ VV^{\top}=I_{p\times p},\ \ \text{and}\ \ \ X=U\Sigma V^{\top},$$
where $I_{n\times n}$ and $I_{p\times p}$ are  identity matrices of dimensions $n$ and $p$ respectively. This is the \textit{singular-value decomposition} (SVD) of $X$. The elements $\sigma_{ii}, 1 \leq i\leq \text{min}(n,p)=r$,  of $\Sigma$ can and shall be chosen to be in decreasing order of $i$.  They are called the \textit{singular-values} of $X$. It is then clear that 
\begin{equation}\label{eq:svd}X=\sum_{i=1}^r \sigma_{ii}u_iv_i^{\top}.
\end{equation}
 This decomposition breaks up the matrix $X$ into simpler components, and all the information about the matrix $X$ are coded in  the singular-values $\{\sigma_{ii}\}$ and the \textit{left and right singular-vectors} $\{u_i\}$ and $\{v_j\}$. The matrices $U$ and $V$ induce \textit{rotation} whereas the matrix $\Sigma$ induces \textit{scaling}. The SVD automatically adjusts for any change in the ordering of rows and columns of $X$. 
 
 The crucial idea is that the singular-values that are large in magnitude are more important, while those that are of lower magnitude could possibly be ignored. If we drop these smaller singular-values and consider a reduced sum on the right side of (\ref{eq:svd}), then it will serve as a good low dimensional approximation for $X$, especially in situations such as ours where $n$ is large. This also helps in computations, and moreover, when $X$ is data dependent, as in our case, the retained singular-values and singular-vectors become important data analytic tools.   

 We use a trimmed version of the SVD in our analysis. We also use generalised SVD, which is a more sophisticated tool and is discussed briefly in Section \ref{sec:suppl_plot}.

\subsubsection{Bergsma's correlation; Spatial Bergsma statistic}\label{sec:bergsma} A value of $0$ for the conventional product-moment correlation (i.e. Pearson correlation) does not imply independence. Many (non-parametric) alternatives try to address this and other issues. One of these is Bergsma's correlation $\rho$ (see \cite{bergsma2005new}), which has the nice property that zero correlation implies independence. This is not a rank based method, like the Bergsma's $\tau$ correlation. For some details about its description, properties and estimation, see \cite{bose}. We chose this correlation since it leads to a nice spatial measure of association that we shall discuss below.   

To investigate spatial association, we need a statistical measure for it. Such a measure is built using two ingredients. Suppose we have $p$ spatial units (for us, $p=280$). 

A \textit{spatial proximity weight matrix} $W=((w_{ij}))_{1\leq i, j \leq p}$ represents the extent of proximity between the spatial locations. A numerical weight $w_{ij}$ is assigned to each pair $(i,j)$, and larger weights signify greater spatial proximity. See  \cite{getis2010analysis} for different popular choices  for $W$. It is always assumed that $w_{ii}=0$ for all $i$. Further, $W$ shall be \textit{row-standardized}, so that each row sum in the matrix is equal to one. We shall use two choices for $W$. The first is the \textit{lag-1 adjacency matrix} 
	\[
	w_{ij}= 
	\begin{cases}
		1& \text{if regions $i$ and $j$ are adjacent locations, } \\
		0              & \text{otherwise.}
	\end{cases}
	\]
The second choice of $W$ is obtained by applying  exponential decay on a notion of distance between the locations. Here simple Euclidean distances between the grids have been considered to construct the base distance matrix.
  
 The second ingredient is a \textit{similarity matrix}. Suppose that we have a series of observations 
 ${X_i} =\{x_{m,i},m=1,\ldots, T\}$ at each of the $p$ locations, $i=1,2,\ldots, p$ for $T$ time points. The similarity matrix $Y$ is defined as $S:=((sim_{ij}))$, where $sim_{ij}$ is some measure of similarity (dependence) between $X_i$ and $X_j$. In the literature on spatial studies, often the chosen measure of similarity is taken to be the product moment correlation. However, our choice for
 $sim(i,j)=\rho(X_i, X_j)$ where $\rho (X_i, X_j)$ is Bergsma's correlation between the variables $X_i$ and $X_j$, mentioned earlier. 

  The \textit{spatial association measure} (\textit{Spatial Bergsma}) $S_B$ proposed by \cite{kappara} uses Bergsma's correlation as follows. Let $X_i$ denote a real variable for the location $i=1,\ldots, p$. Then $S_B$ with a row-standardized $W$ is defined as 
\begin{eqnarray}
	S_{B}
 &=& p^{-1}\sum_{1\leq i <j \leq p}(w_{ij}+w_{ji})\rho(X_i, X_j).\label{SBp1}
\end{eqnarray}

Suppose that we have observation vectors ${X_i}=\{x_{m,i},m=1,\ldots, T\}.$ Let  $\tilde{\rho}^{(ij)}$ denote the estimated $\rho$, calculated for the pair $(X_i,X_j)$ (see \cite{bose} for details). Then an estimate of $S_{B}$ is the spatial Bergsma's statistics, given by 
\begin{equation}
	\tilde{S}_{B}:=p^{-1}\displaystyle{\sum_{1\leq i <j \leq p}(w_{ij}+w_{ji})\tilde{\rho}^{(ij)}},\label{sb}
\end{equation} 
The {\tt R} code in \cite{bose} (Appendix) can be used for computing $\tilde{S}_{B}$. Computation of  $\{\tilde{\rho}^{(ij)}\}$ involves computation of several $U$-statistics with complicated kernels. When the number of observations is large, this is computationally very intensive. Thus, we use this concept only on the yearly average data.

The asymptotic distribution of $\tilde S_B$, as the number of observations becomes large, is known only when the observations are independent and identically distributed across time. This may appear to be a major hurdle since the DTR values are not expected to show any sort of independence. Nevertheless, this can still be used fruitfully as a comparative and a diagnostic tool once the data is reasonably time-detrended as we have done in our analyses.

\section{Preliminary exploration}\label{sec:preliminary_exploration} Our primary focus of study is spatial association, and temporal changes therein, in the DTR data arising from India. As mentioned earlier, complete data is available only for 280 grid points out of 362, and most of our analysis is restricted to these. However, for certain statistics involving averages, we do use the partial data available for the additional 82 grids points.

\subsection{Empirical spatial distributions of average DTR} We first focus on the empirical spatial distribution of average DTR over various temporal subsets. For this part we use data from all the 362 grids. The data can be grouped in various ways, for example, year-wise, or season-wise within each year. The locations can be stratified according to climatic regions too.
\vskip5pt
\begin{figure}[!ht]
     \centering
     \includegraphics[width=0.8\linewidth,height=2in]{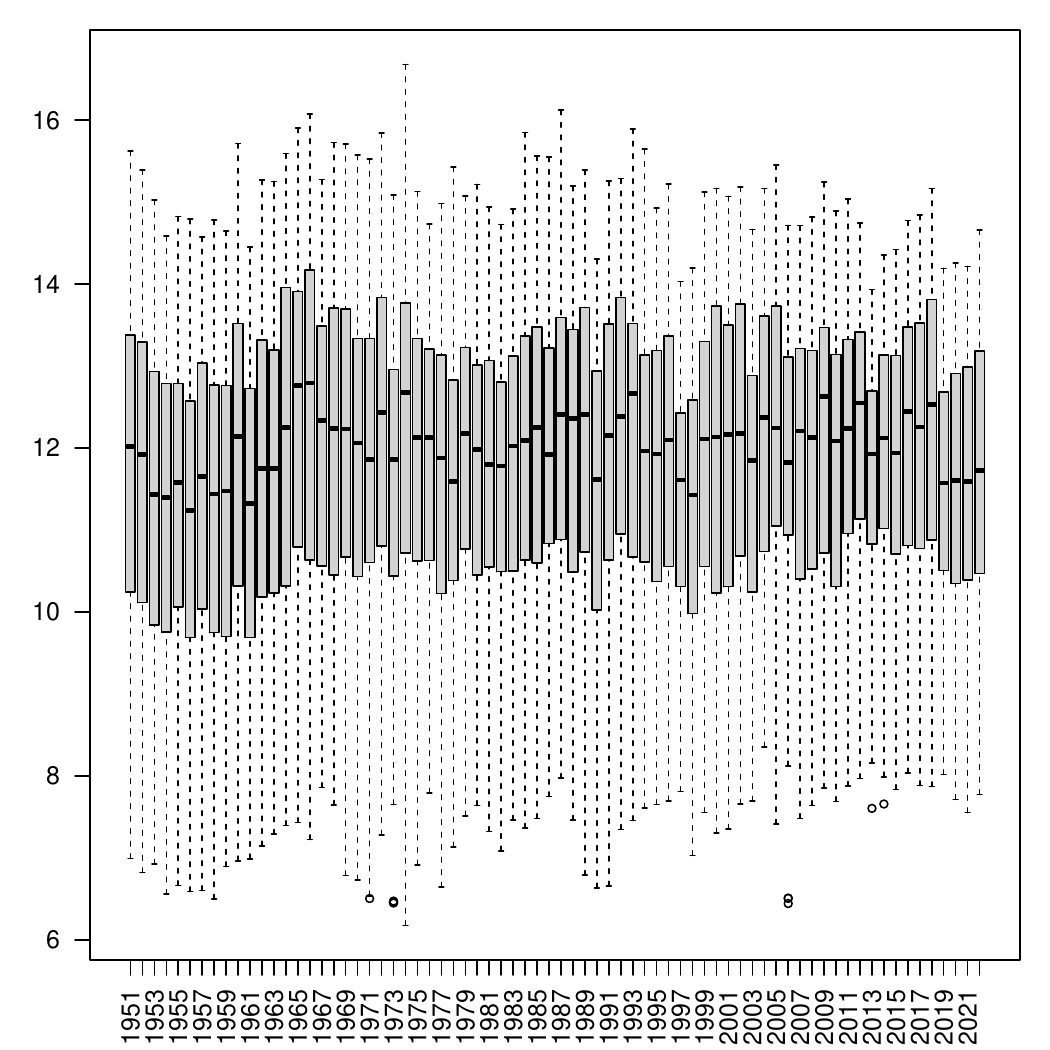}\ \ 
     \caption{Yearly average DTR values (1951--2020) for the 362 grids.}
         \label{fig:362grid_dtr_yearly_average}
\end{figure}

In Figure \ref{fig:362grid_dtr_yearly_average}, for each year on the horizontal axis, we have plotted the yearly average DTR values from the 362 locations on the vertical axis. Visually, the dispersion of the underlying average DTR distribution has generally decreased over the years, and there is visible shrinkage of distribution from both sides. There is also an indication of change in the pattern somewhere in the 1960’s. 
\vskip5pt
\begin{figure}[ht!]
     \centering
     \includegraphics[scale=1, height=4in]{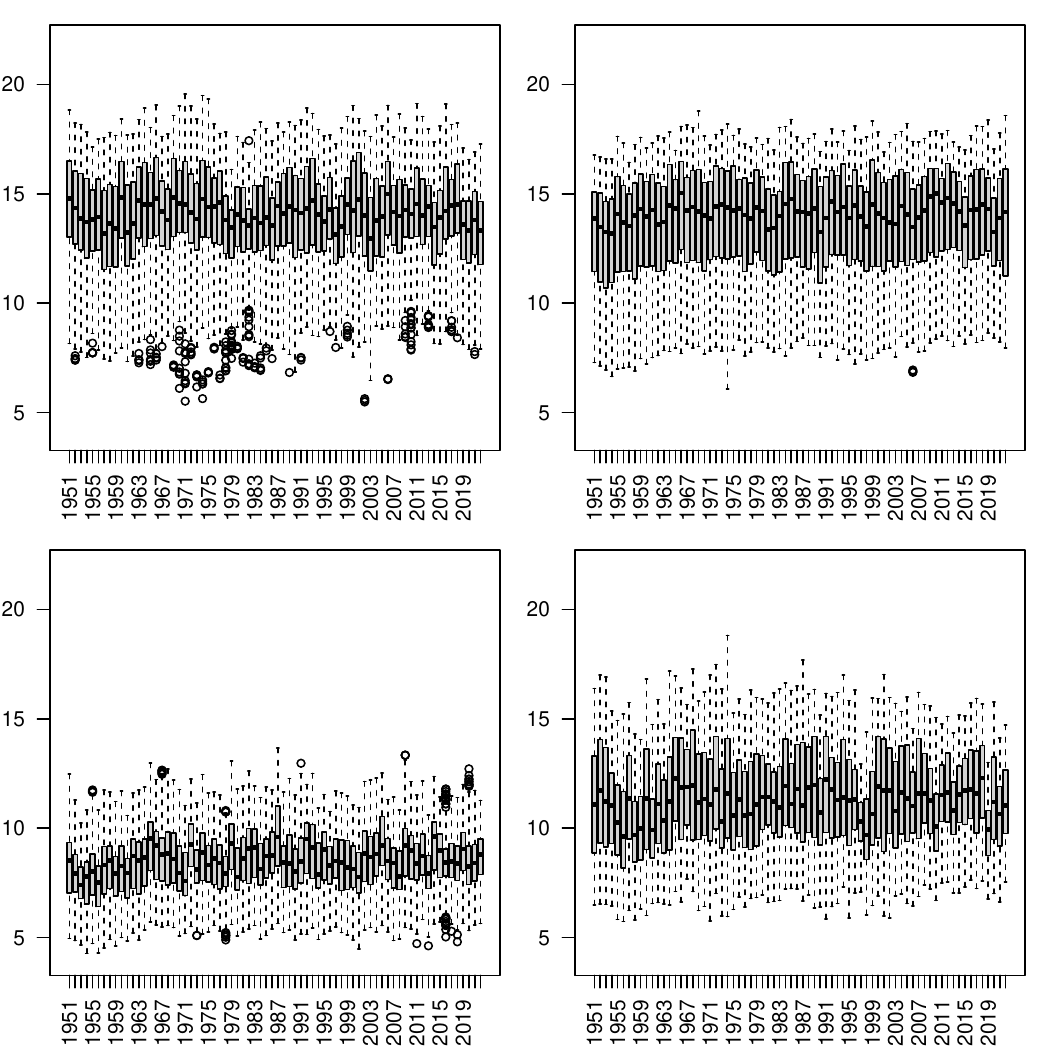}\ \ 
     \caption{Seasonal average DTR values (1951--2020) for the 362 locations. In \textit{clock-wise direction}; DJF (Dec.--Feb.), MAM (March--May), SON (Sept.--Nov), JJA (June--Aug.).}
    \label{fig:362grid_seasonal_average}
\end{figure}
The yearly weather of India is generally divided into four seasons: December to February (DJF), March to May (MAM), June to August (JJA), and September to November (SON). In Figure \ref{fig:362grid_seasonal_average}, for each of the 72 years on the horizontal axis, we have plotted the DTR averages of each season (\textit{seasonal averages}) for the 372 grids on the vertical axis. The behaviour of the DTR varies across seasons. The shrinkage of the distribution that was visible 
in Figure \ref{fig:362grid_dtr_yearly_average} is particularly pronounced during autumn (SON), and to a lesser extent in winter (DJF). The spatial spread of DTR is narrowest and most homogeneous during monsoon (JJA) and is widely varying in the other three seasons.

The average DTR values could also be plotted after stratifying the grids by climatic zones or by climatic zones and seasons. Some such plots are given in Section \ref{sec:suppl_plot}.

\subsection{Empirical spectral distribution of DTR correlation matrices} 
\begin{figure}
\centering
\begin{subfigure}{.496\textwidth}
  \centering
    \includegraphics[width=0.8\linewidth, height=2in]{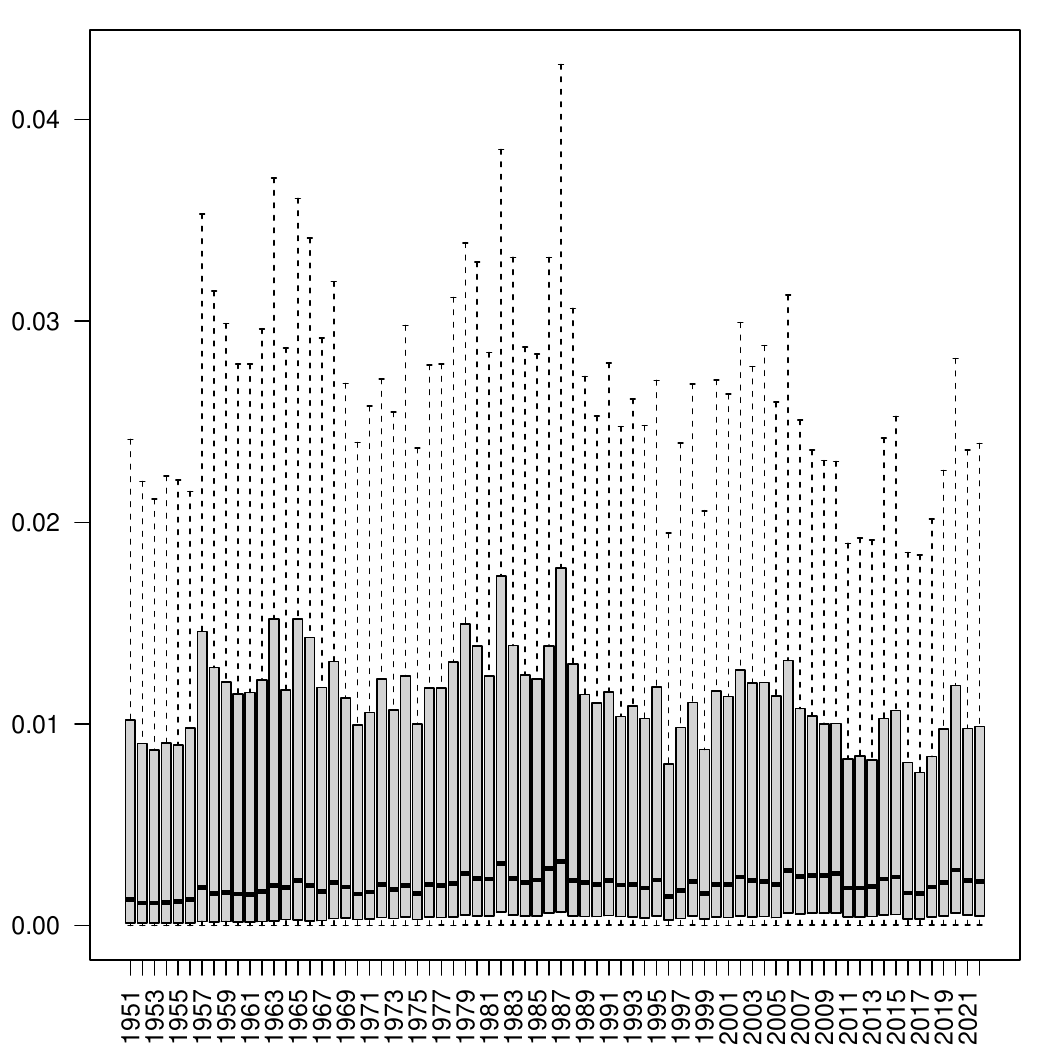}
  \caption{ESD of $R_i^{D}$, $1\leq m \leq 72$. }
    \label{fig:ESD_of_cor_mat_of_org_DTR}
\end{subfigure}
\begin{subfigure}{.496\textwidth}
  \centering
    \includegraphics[width=0.8\linewidth, height=2in]{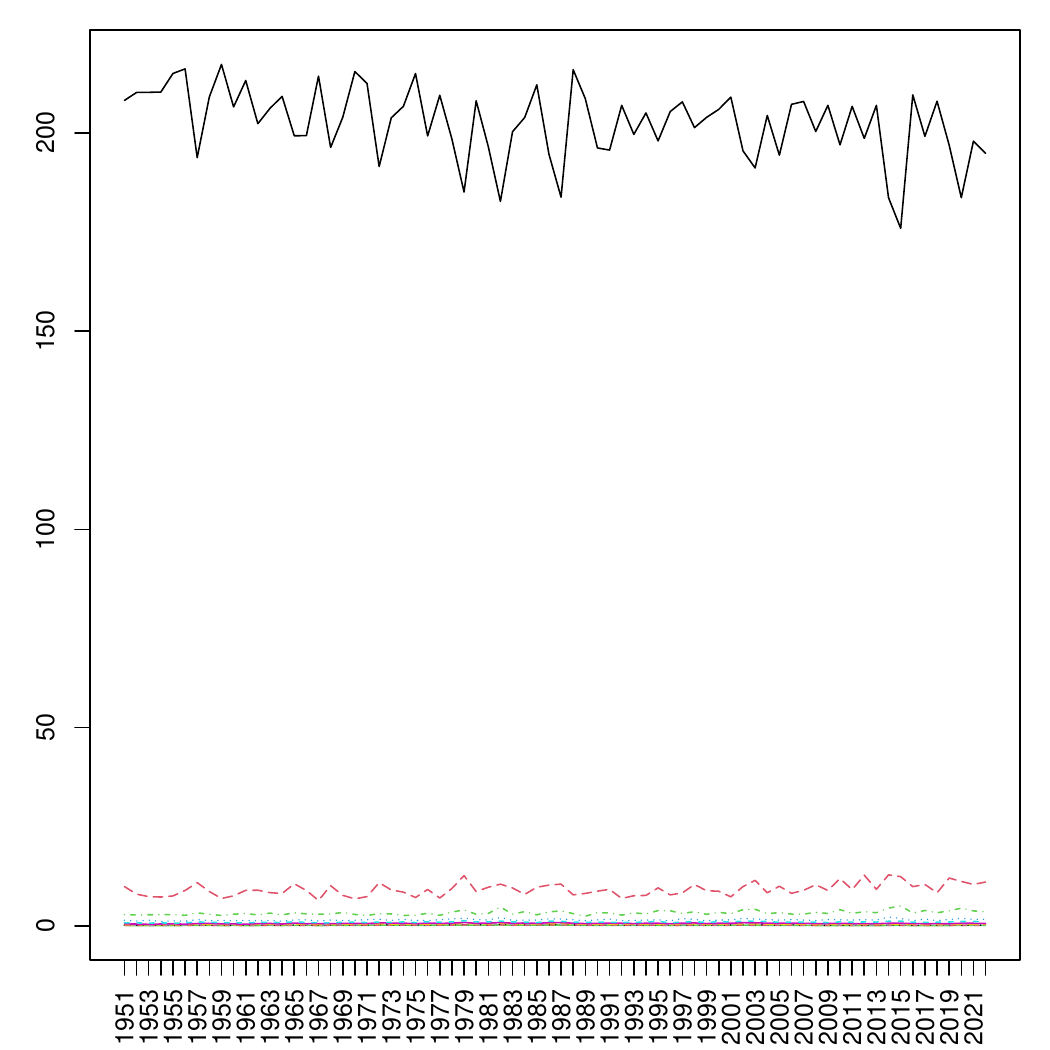}
    \caption{Quantiles of the eigenvalues for each year.}
  \label{fig:Top_eigenvalues_of_cor_mat_of_org_DTR}
\end{subfigure}
\caption{ESD analyses of the 72 correlation matrices $\{R^{D}_i\}$.}
\label{fig:Eigenvalues_of_yrly_cor_mat_of_org_DTR}
\end{figure}
We now consider the (product-moment) correlation matrices $\{R^D_i\}$ obtained from the daily DTR values for all pairs of the 280 grid points, for every year $i$. We shall study these random matrices to assess the changes in the underlying spatial association pattern over the years, as has been suggested by Figures \ref{fig:362grid_dtr_yearly_average} and \ref{fig:362grid_seasonal_average}.
\vskip5pt

\noindent In Figure \ref{fig:ESD_of_cor_mat_of_org_DTR} we have presented the ESD for each year, based on $R^D_i$. 
Over the years there has been a small but steady increase in the median value of the ESDs, and a slight shift away from zero in the left tail of the distribution. Since the sum of eigenvalues is fixed for a correlation matrix, the above, generally speaking, will mean that the right tail shrinks towards the middle too. In Figure \ref{fig:Top_eigenvalues_of_cor_mat_of_org_DTR}, we have plotted the quantiles (at 10\% interval) of the eigenvalues for each year. We notice a small but gradual decrease in the top eigenvalues, although noticeable oscillation in these values is also present.

\subsection{Correlation matrix \texorpdfstring{$R^D$}{xxx} and its de-noised version \texorpdfstring{$\hat{R}^{D}$}{xxx}}
\begin{figure}[ht!]
    \centering
    \includegraphics[width=0.5\linewidth]{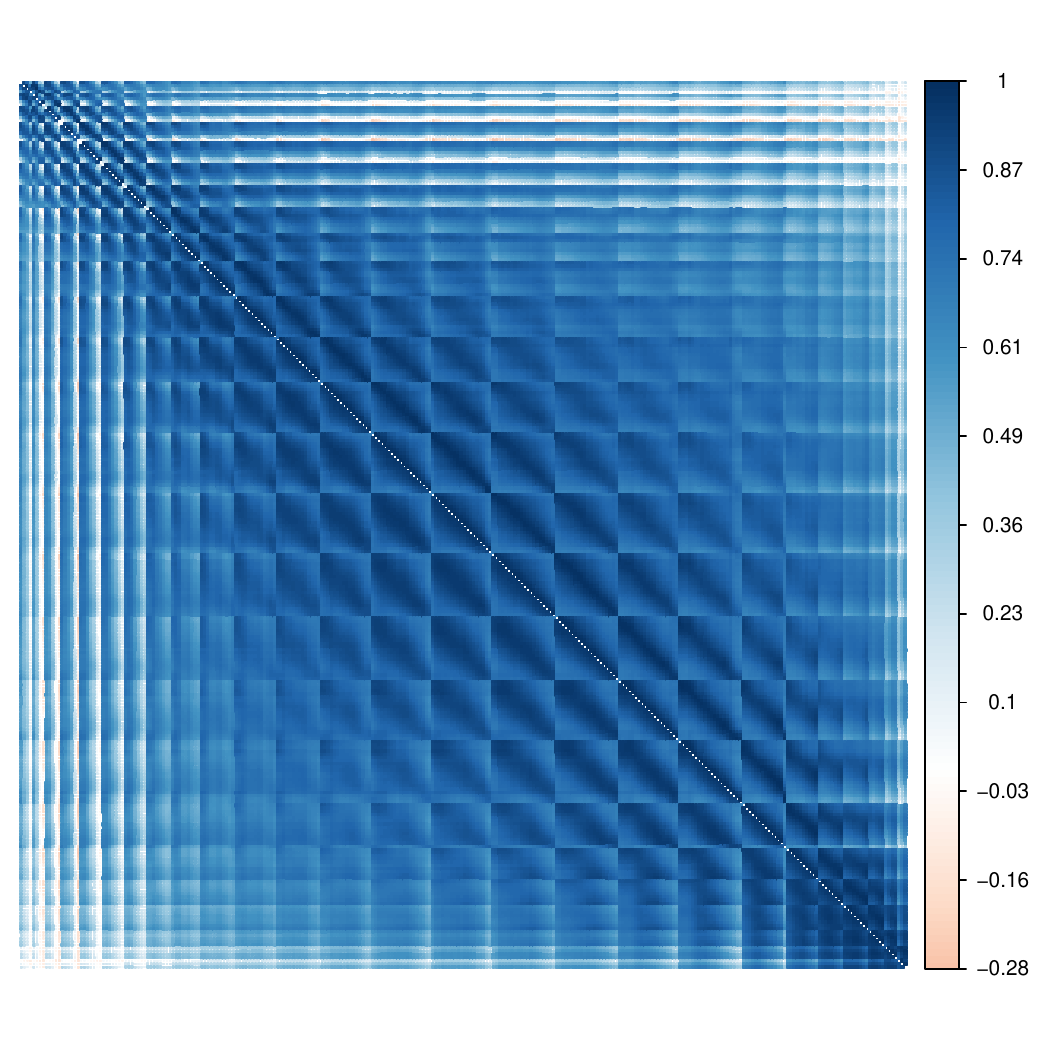}
    \caption{\textit{Upper triangle}--entries from the correlation matrix $R^D$; \textit{Lower triangle}--corresponding entries from the MP de-noised version of $R^D$.}
    \label{fig:DTR_cor_MPdns}
\end{figure}

Let $\{\lambda^{D}_i\}$ be the eigenvalues of $R^{D}$. Applying the MP law based cutoff, only $10$ of these were significant. 
Then $R^{D}$ is approximated by using these eigenvalues and the corresponding eigenvectors $\{e^{D}_i\}$ as the 
the MP law based \textit{de-noised matrix} $\hat{R}^{D}$: 
$$\hat{R}^{D} = \sum_{i=j}^{10} \lambda^{D}_{j}  e^{D}_{j} {(e^{D}_{j})}^{\top}.$$
The upper and lower triangles in Figure \ref{fig:DTR_cor_MPdns} are based on $R^D$ and $\hat{R}^{D}$ respectively. First note that from the upper triangle, there is visible association between spatial locations. Indeed, of the $10$ top eigenvalues, the largest is ten times in magnitude compared to the second largest, as is evident  from Figure \ref{fig:Top_eigenvalues_of_cor_mat_of_org_DTR}. This may not be surprising since common temporal pattern arising due to seasonality would naturally lead to a strong association/correlation. 
Nevertheless, presence of high degree of positive association might be masking other patterns. MP law based de-noising has some effect and brings balance to the distribution of the correlations (observe the negative values along the left vertical). However, the correlation distribution still looks heavily positive. 
\vskip5pt
Thus although the MP law has been very useful in many applications, it is not effective here in bringing out any pattern, when implemented on $R^{D}$.


\section{Analysis}\label{sec:analysis} The preceding explorations indicate that in order to bring out meaningful spatial association patterns, it is imperative to remove association due to predominant temporal patterns. Time series decomposition was carried out for each of the 280 daily-DTR-time series. Time trends and seasonal components were removed and the spatial correlation matrix $R^{T}$ was obtained based on the resulting time-detrended data $T$. By applying MP law, 
there were $18$ significant eigenvalues and once again there is only one dominant eigenvalue. Thus we explore alternate methods for removing dominant trends to bring out the spatial association pattern underlying the DTR data.

\subsection{Trimmed-DTR matrix \texorpdfstring{$S$}{xxx}: detrending by SVD}\label{sec:trimming_explained} SVD is used to project the data onto a lower ranked subspace. The common strategy is to use the truncated SVD after dropping the relatively smaller singular-values. 

However the objective here is to remove the predominant patterns to elucidate the finer and yet unknown (spatial) patterns. For $D$, the presence of a few (predominantly temporal) strong signals are masking other (spatially) significant signals. Our strategy would be to remove components corresponding to a few dominant singular-values of $D$, and work with the resulting \textit{trimmed DTR matrix} $S$.  

\begin{figure}
    \centering
    \includegraphics[width=0.35\linewidth]{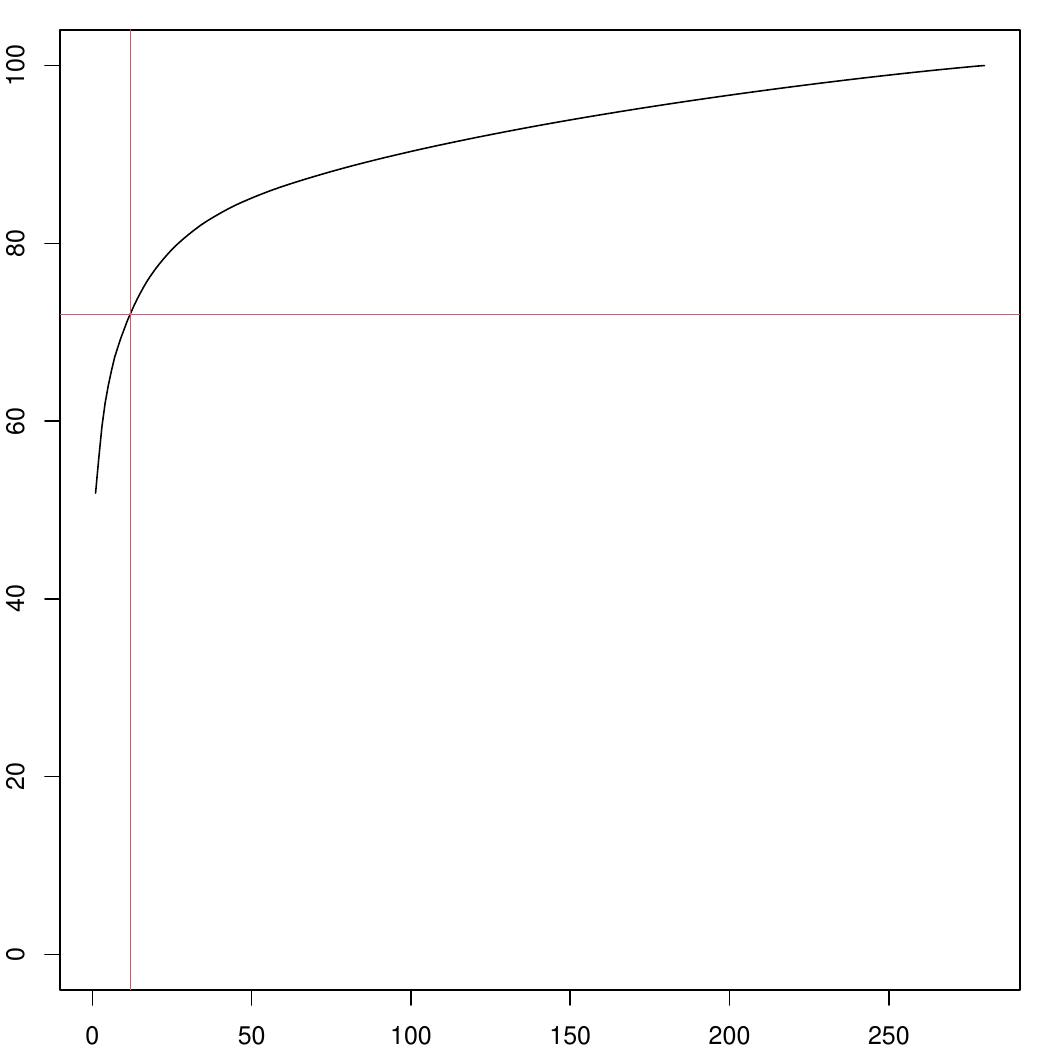}
    \caption{Cumulative singular-values for the original DTR data $D$.}
    \label{fig:Singular_values_of_org_DTR}
\end{figure}

\begin{figure}
    \centering
    \includegraphics[width=0.35\linewidth]{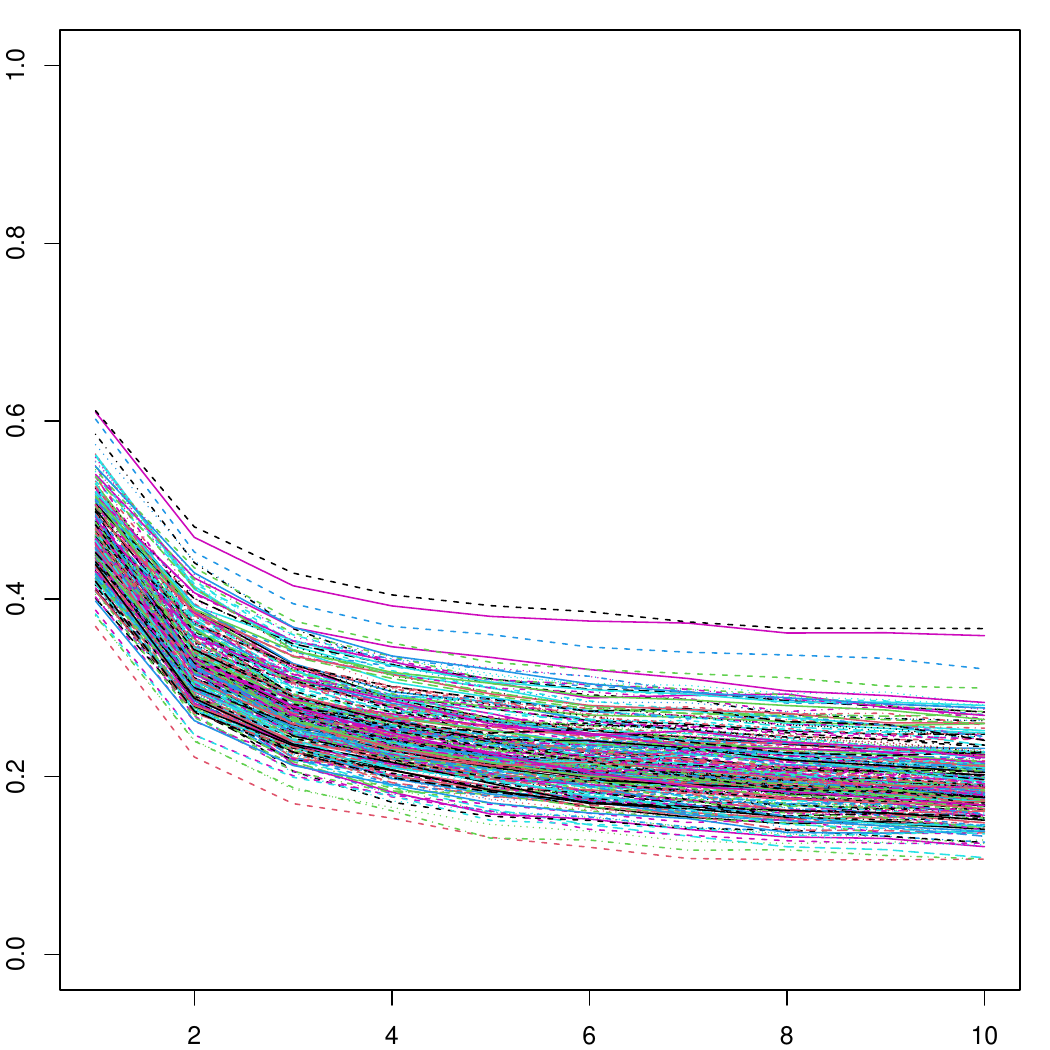}
    \caption{Autocorrelation series for the 280 time series based on trimmed data $S$.}
    \label{fig:ACF_of_org_DTR_data}
\end{figure}

For that purpose we identified the minimal number of singular-values $D$ required to reasonably \textit{time-detrend} the data, while retaining as much of the overall signal as possible. 
Efficacy of time detrending was assessed by computing the \textit{autocorrelation function} (ACF), and the overall signal coverage was measured by the \textit{percent share of singular-values} retained. Employing these two criteria, we stopped after the top 12 singular-values. Referring to Figure \ref{fig:Singular_values_of_org_DTR}, these cover 72\% of the singular-values. Taking out the components $\sigma_{ii} u_iv_i^{\top}$ from $D$, $1\leq i \leq 12$, we obtain 
the \textit{trimmed data matrix} 
\begin{equation}S=D-\sum_{i=1}^{12}\sigma_{ii} u_iv_i^{\top}.
\label{eqn:S}\end{equation}
The autocorrelation plots of the time series from each row of $S$ given in Figure \ref{fig:ACF_of_org_DTR_data} show that this trimming removed a lot of temporal patterns.


\begin{figure}
    \centering
    \includegraphics[width=0.8\linewidth, height=0.5\linewidth]{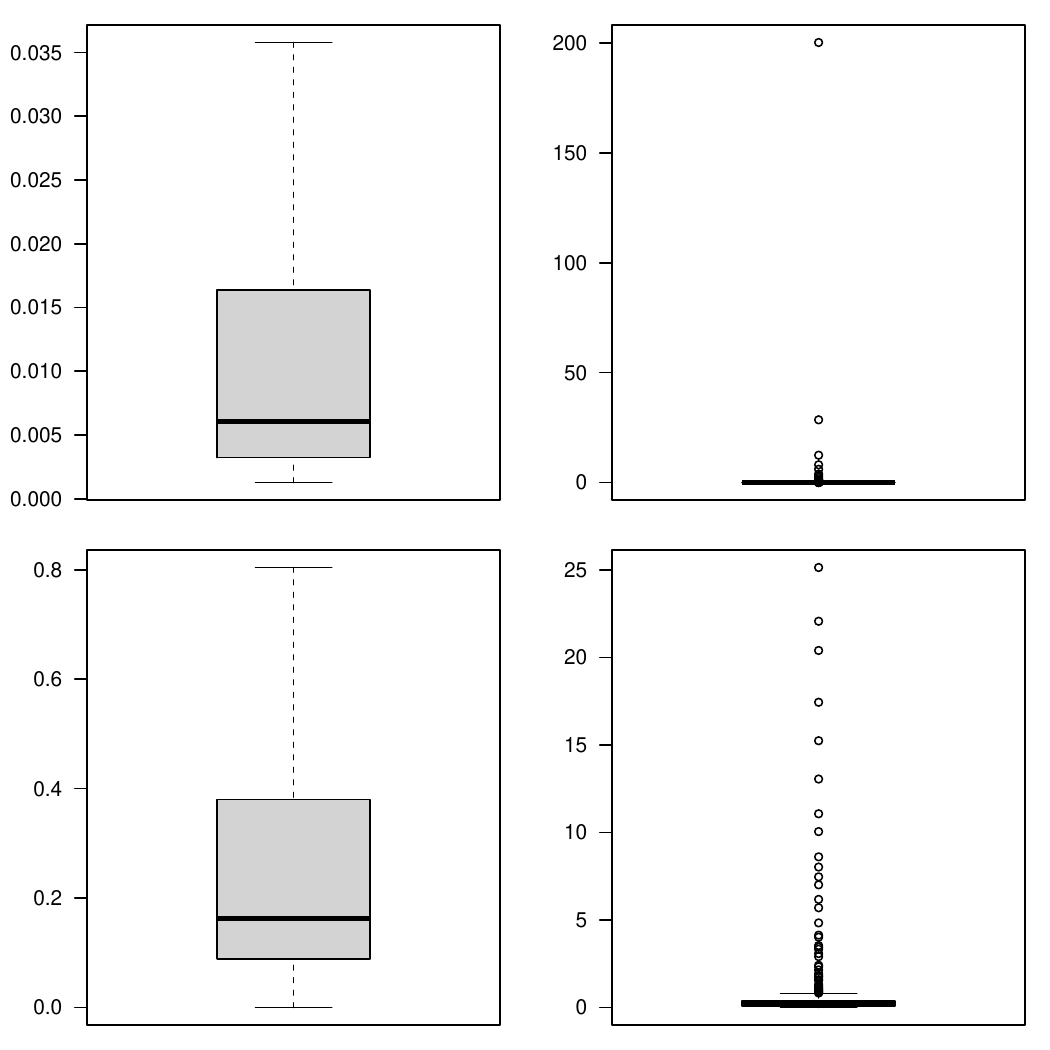}
    \caption{Eigenvalues of the correlation matrices: \textit{top panel}, $R^{D}$;
        \textit{bottom panel}, $R^{S}$.}
           \label{fig:ESD_org_DTR_and_SVR10res}
\end{figure}
Now consider the correlation matrix $R^S$ obtained from $S$. Applying the MP law on the ESD of $R^S$ now shows $33$ significant eigenvalues, compared to only $10$ for $R^D$ (and $18$ for the correlation matrix based on the residuals from time-series decomposition $T$, as mentioned  earlier). See Figure \ref{fig:ESDs} on eigenvalues of correlation matrices in Section \ref{sec:suppl_plot}.

In addition to these, we also consider the entire ESDs. Figure \ref{fig:ESD_org_DTR_and_SVR10res} shows that the ESD of $R^{D}$ is extremely long tailed. The ESD for $R^{S}$ has a much more balanced distribution.
Trimming has shifted the spatial association behaviour, as captured by the yearly correlation matrices, and also made the distributions more balanced. Later in Section \ref{sec:chng_pt} we have also considered the behaviour of the spectral distribution of the yearly correlation matrices $\{R^S_i\}$. In Section \ref{sec:suppl_plot} we have given further information on the singular-vectors removed. 

\subsection{Generalised SVD to assess spatio-temporal information in \texorpdfstring{$S$}{xxx}} 
\begin{figure}
\centering
\begin{subfigure}{.45\textwidth}
  \centering
    \includegraphics[width=1\linewidth]{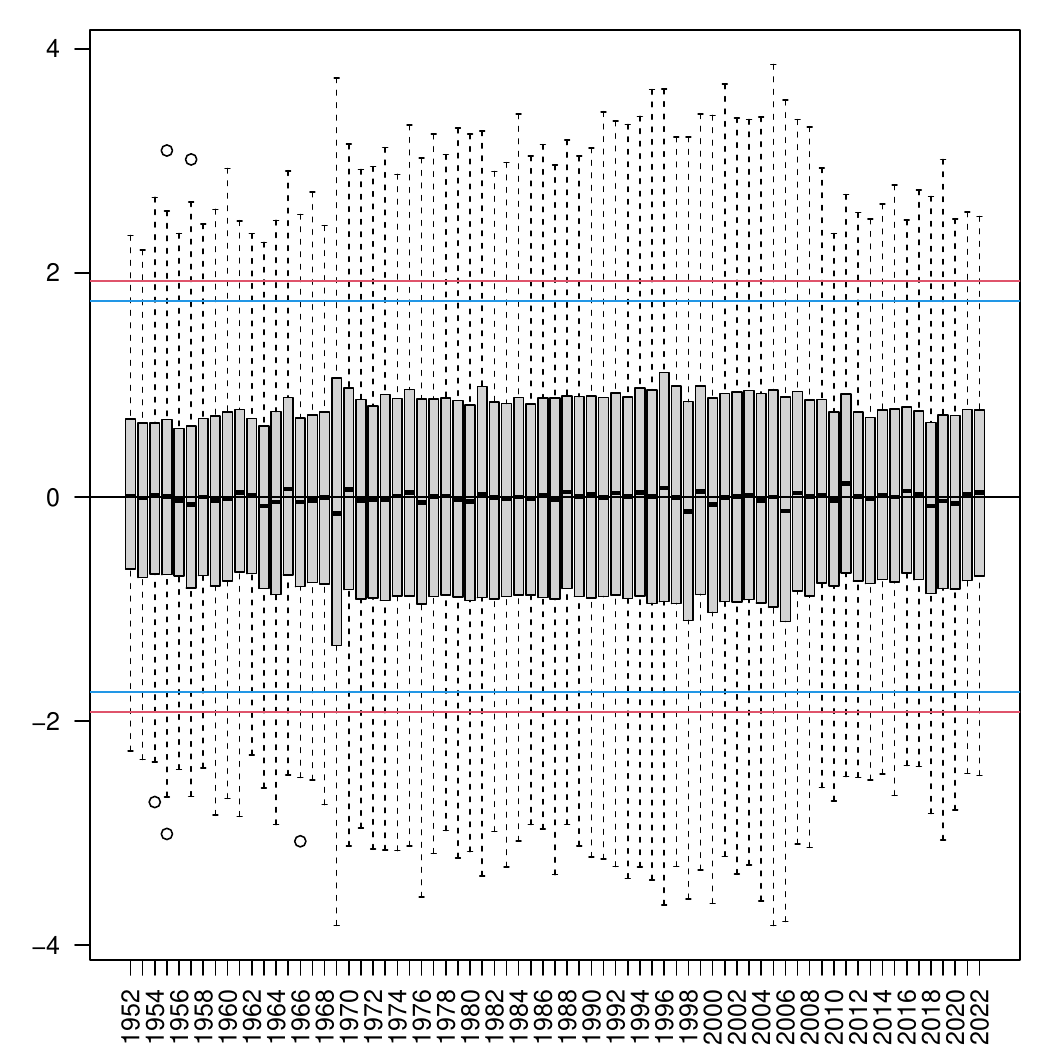}
    \caption{Original DTR data $D$.}
  \label{fig:GSVD_yrly_org_DTR}
\end{subfigure}
\begin{subfigure}{.45\textwidth}
  \centering
    \includegraphics[width=1\linewidth]{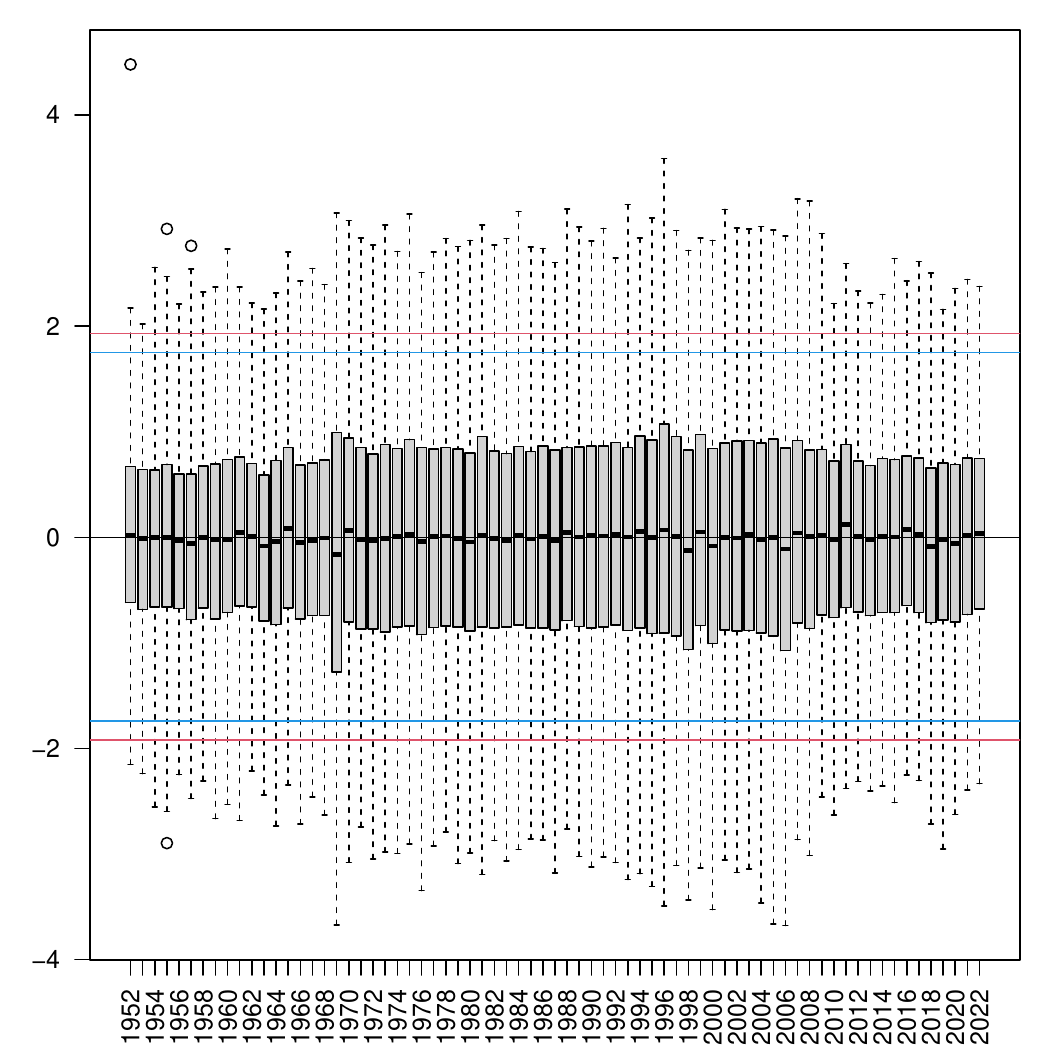}
  \caption{Trimmed DTR data $S$.}
  \label{fig:GSVD_yrly_SVD10s_DTR}
\end{subfigure}
\caption{Empirical distributions of the (log) generalised singular-values based on 
pairwise successive years data from $D$ (left panel) and $S$ (right panel). The (empirical) critical values for the generalised SVs are marked by horizontal lines.}
\label{fig:GSVDs}
\end{figure}

The method employed above is not a traditional de-trending technique. We need to assess if there has been loss of spatial information. We use GSVD for this purpose. See Section \ref{sec:GSVD} for  a brief description of GSVD. For each successive pairs of years data from both $D$ and $S$ (resulting in 71 such pairs from each), we computed the GSVD. The results are shown in Figure  \ref{fig:GSVDs}, where we have given the boxplots for the log-generalised singular-values for all such successive pairs, separately for $D$ and $S$.

Recall that in GSVD the right singular-vectors are held common while the left singular-vectors are allowed to vary between the two matrices. In the current context this translates into holding the vectors capturing the spatial rotation fixed, while allowing the temporal rotation to vary from year to year. 

As Figure \ref{fig:GSVDs} shows, the empirical distributions of the generalised singular-values remain same or similar for pairs arising from $D$ and $S$. This confirms that the temporal information retained in $S$ is comparable to that in $D$.

Recall that to identify significant signals, the  MP law was used for the eigenvalues of the correlation matrices. To carry out similar exercise for the generalised singular values, we needed a similar result on the null distribution of generalised singular values. This seems to be missing in the literature.

Hence we estimated the null distribution of GSVs empirically (see Section \ref{sec:null-GSVD} and Figure \ref{sec:suppl_plot}) and obtained the critical values from that. By comparing the observed GSV distributions and the critical values, it appears that (in the log scale) the GSVs for $S$ has a more ``regular" distributional behaviour compared to that for the original DTR matrix.

We have also carried out the reverse exercise to assess spatial information retention using  $D^{\top}$ and $S^{\top}$ (see Section \ref{sec:sp-GSVD}) with similar conclusion as above. 

This exercise assures that $S$ can be used as proxy for ``time-detrended” series, and it retains the spatial features of $D$. Hence $S$ is suitable for studying potential spatial patterns in the DTR data, and will be considered next for further analyses.

\subsection{Exploring spatial association pattern} 
Now we consider the correlation matrix $R^S$ obtained from $S$. Prior to ascertaining existence of any pattern within that matrix, we first explored it from various perspectives.

\begin{figure}
    \centering
    \includegraphics[width=0.8\linewidth]{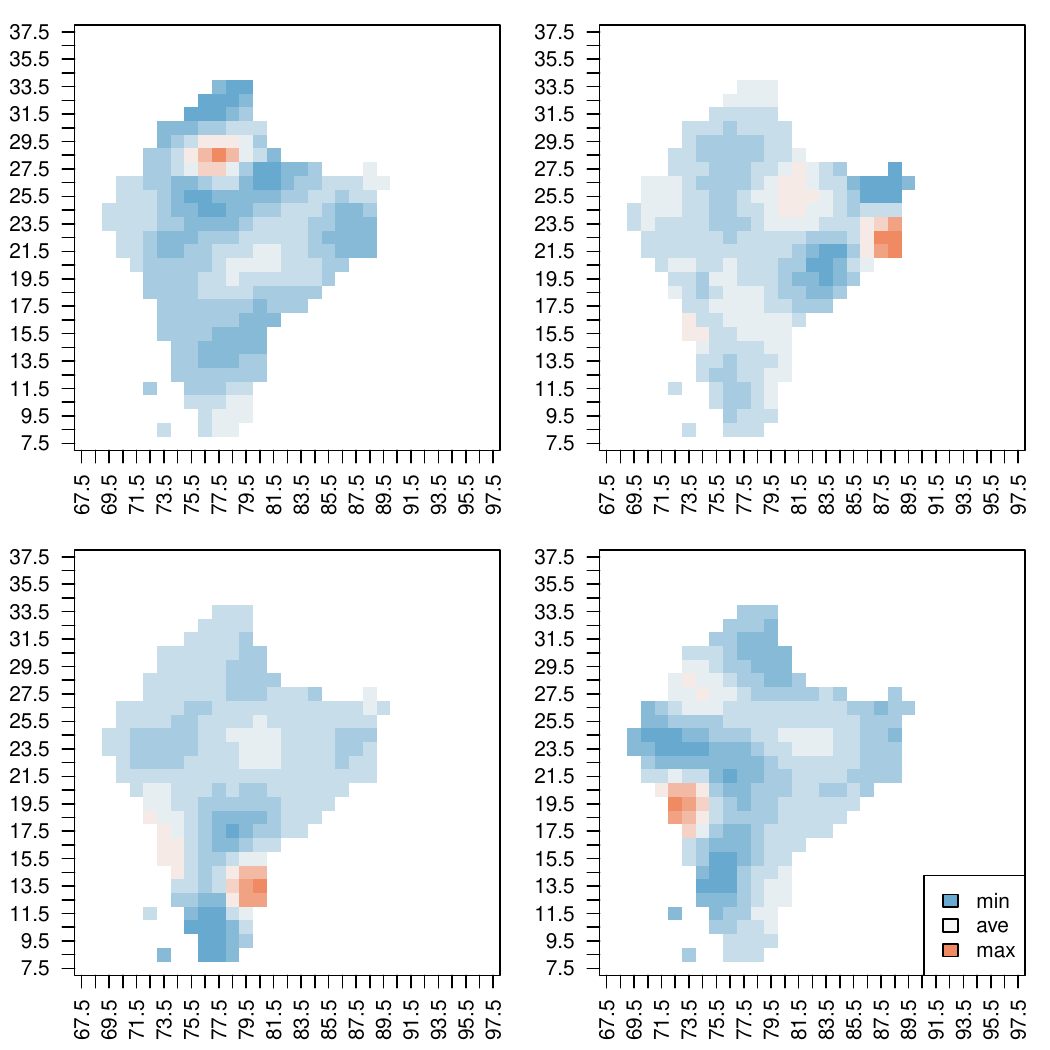}
    \caption{Correlations within grids for $S$ that cover each of the four major cities. \textit{top left}, Delhi; \textit{top right},  Kolkata; \textit{bottom left}, Chennai; \textit{bottom right}, Mumbai.}
    \label{fig:IMD_DTR_SVD12s_res_corr_4cities.pdf}
\end{figure}

For example we identified a few selective grids that cover some of the major cities in India, and plotted the correlation vectors in the form of heat maps corresponding to the latitude and longitude information of individual grids. The resulting images given in Figure \ref{fig:IMD_DTR_SVD12s_res_corr_4cities.pdf} exhibit strong spatial patterns around these cities.
There appears to be strong association in the immediate vicinity for each of these cities. 
 
\begin{figure}
    \centering
    \includegraphics[width=0.5\linewidth, height=2in]{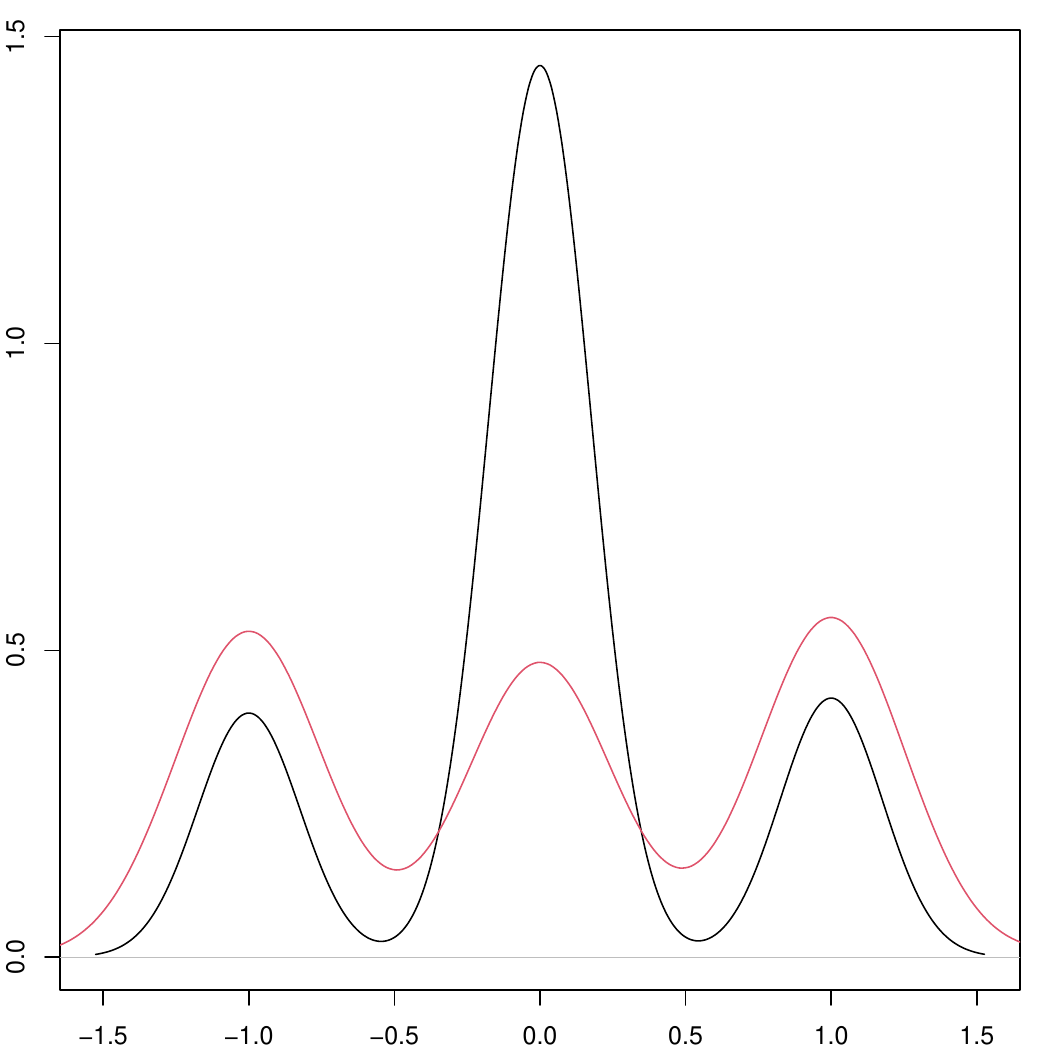}
    \caption{Distribution of distance between grids with maximum association, measured in terms of differences in latitude (in black) and longitude (in red) for $S$.}
    \label{fig:Max_cor_lat_long}
\end{figure}

\begin{figure}
    \centering
    \includegraphics[width=0.5\linewidth]{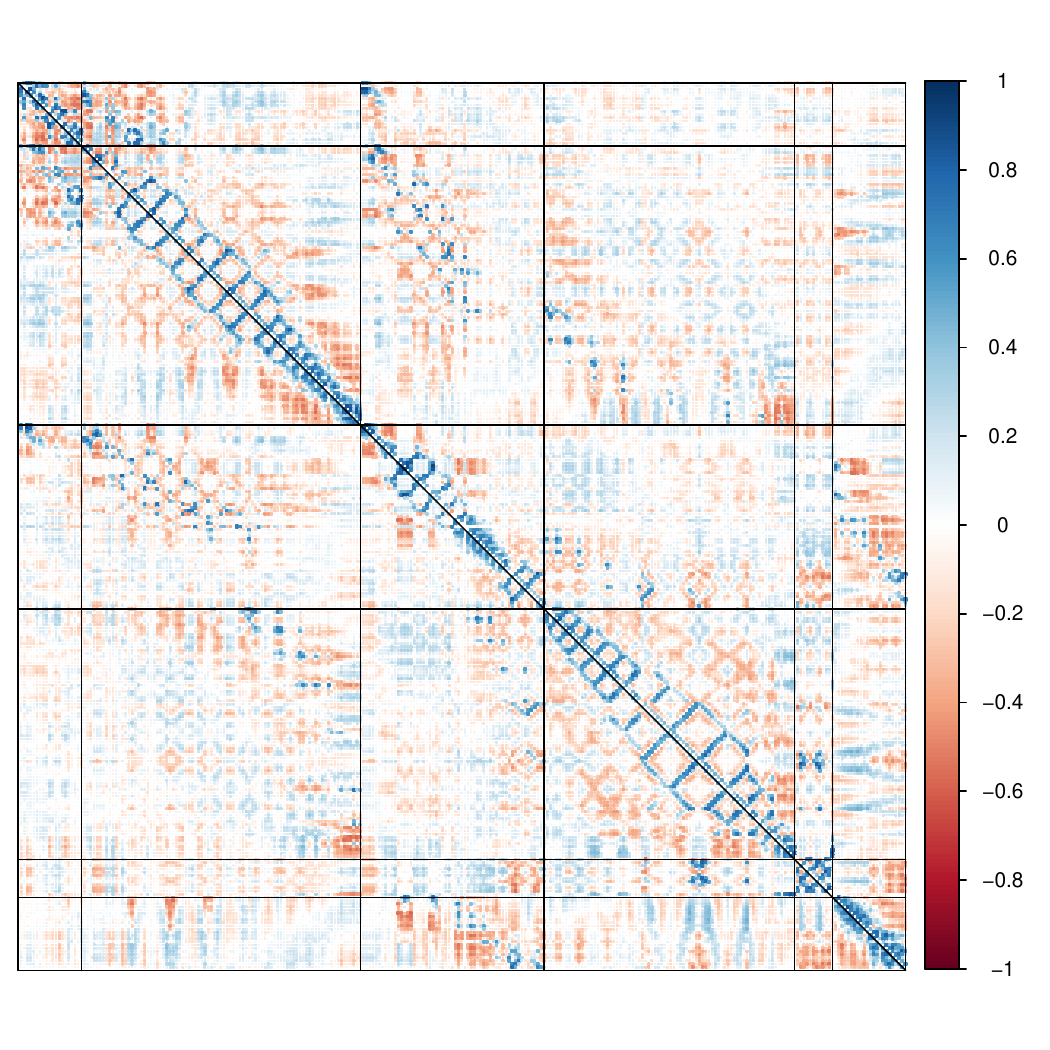}
    \caption{Correlations based on $S$ with grids arranged first according to climatic regions and then in spiral Hilbert space filling curve manner. \textit{Upper triangle}, original correlation matrix; \textit{Lower triangle}, the MP de-noised version.}
    \label{fig:Spiral_ord_clmtc_reg_corr_mat_with_MPdns}
\end{figure}

Let us fix a grid $g_1$, and consider the grid $g_2$ which yields the maximum correlation with $g_1$. Consider the difference in latitudes and longitudes between  $g_1$ and $g_2$. The distribution of these differences across all grids has a significant mode at $0$, while the distribution of the difference in longitudes is approximately uniform around $\pm 1$. In Figure \ref{fig:Max_cor_lat_long}) we have plotted these two distributions. Such salient patterns were not visible earlier but are emerging now.
The question remains whether we can present this pattern for all grids via a single plot. We first stratify the grids according to the climatic regions they belong to. Following which we employ the Hilbert space filling curve approach and arrange the grids in a spiral manner described earlier. We present in Figure \ref{fig:Spiral_ord_clmtc_reg_corr_mat_with_MPdns} the correlation matrix with the grids ordered as described above.
This enables us to visualize much more clearly the spatial association pattern, and also indicates variation in strength and range where maximum association between locations take place. Analysis of the eigenvectors corresponding to significant eigenvalues as per the MP law support the observation of climatic region-specific spatial association pattern.

\subsection{Exploring temporal changes in spatial association pattern}\label{sec:chng_pt} 
The inherent assumption in obtaining the preceding pattern is that it is temporally static, and hence the entire data was used to arrive at the conclusions. To study similar association patterns for any smaller time interval, for example individual calendar year, would result in the task of summarising numerous such matrices.

\begin{figure}
    \centering
    \includegraphics[width=0.5\linewidth, height=2in]{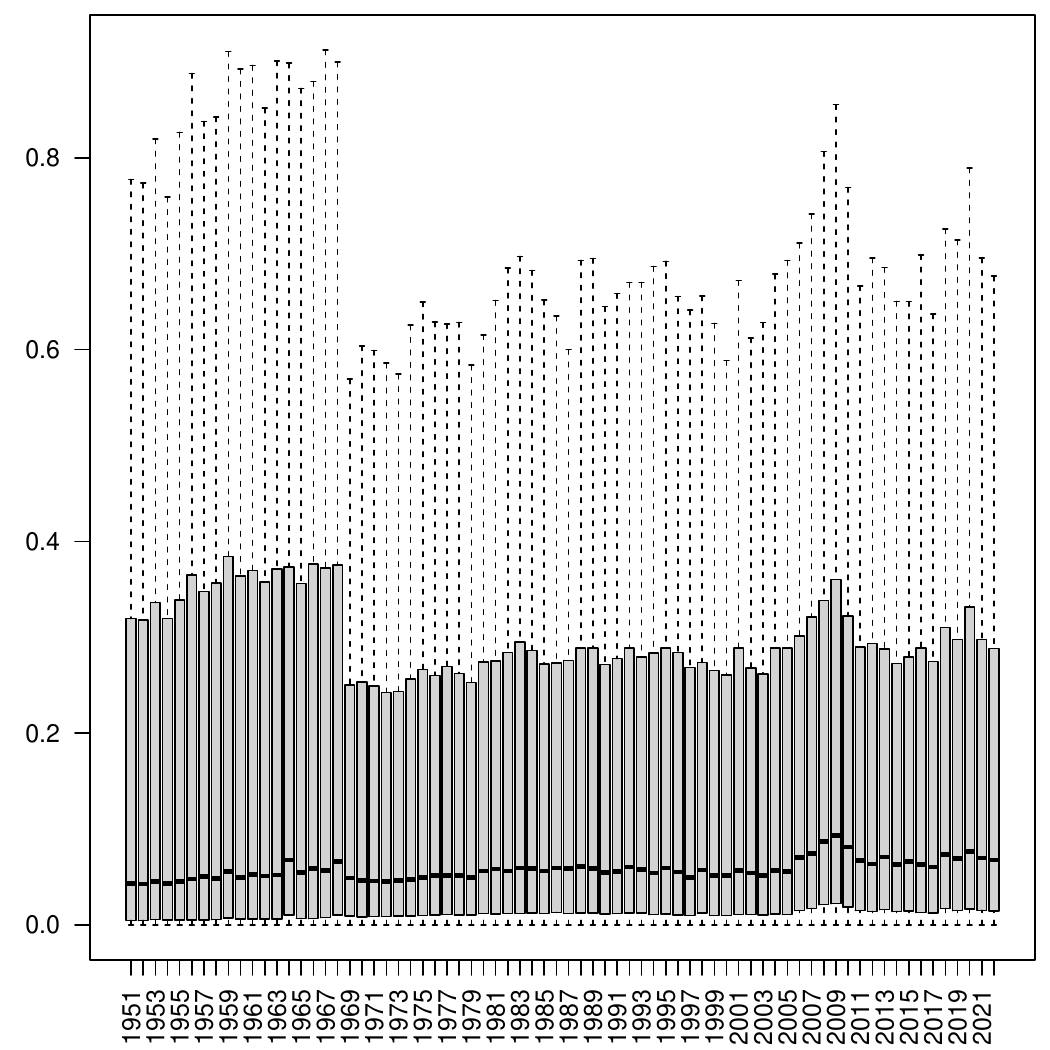}
    \caption{Eigenvalues of $R^{S}_i$.}
        \label{fig:Bulk_ESD_yrly_cor_mat}
\end{figure}

We achieve this by studying the ESD of each of the correlation matrices $\{R^{S}_i\}$. Studying the ESD in lieu of the entire correlation matrix enables us to investigate for patterns in a much smaller dimension while retaining the essential features of the data.
Figure \ref{fig:Bulk_ESD_yrly_cor_mat} presents the ESD for each year.
The ESDs clearly indicate the occurrence of a drastic change in the spatial association pattern in the late 1960s.

\subsection{Spatial Bergsma: A univariate spatial association measure} 
The correlation matrices present a $280\times 280$ dimensional summary of the association, and the ESDs present a $280\times 1$ dimensional summary. We use spatial Bergsma measure of \cite{kappara} to arrive at a univariate measure for the spatial pattern observed. As described earlier in Section \ref{sec:bergsma}, the key ingredients to derive such a measure are a spatial weight matrix and a similarity measure. We used two candidates for spatial weight matrix, namely the lag-1 adjacency matrix and the exponentially decaying distance matrix. For similarity measure we would be using Bergsma's $\rho$ for its useful ability to depict independence (corresponding to the zero value of the measure). 

\begin{figure}
\centering
  \begin{subfigure}{.31\textwidth}
      \subfloat[a][Yearly all India data]{\includegraphics[width=1\linewidth]{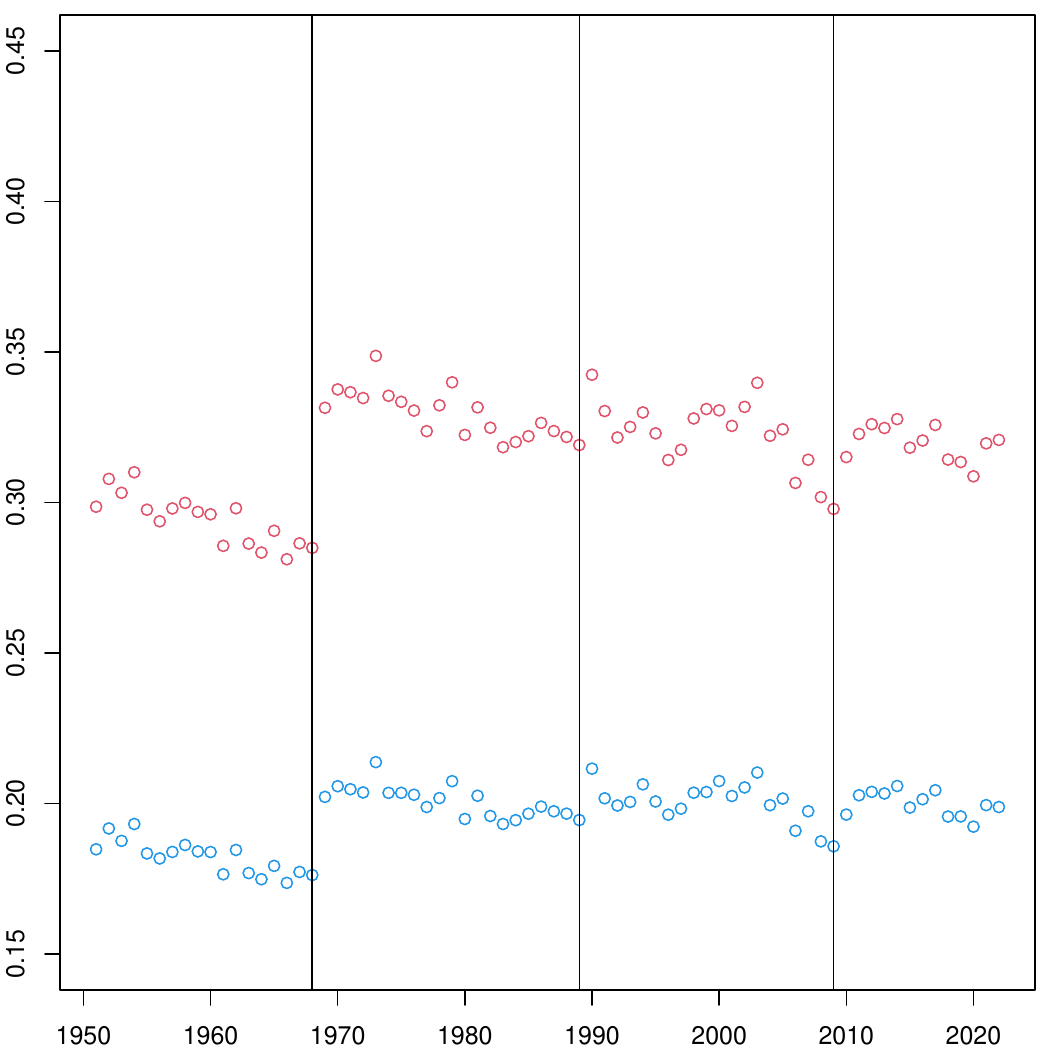}
      \label{fig:SB_yrly_SVD10s_DTR}} \\
      \subfloat[b][Monthly all India data]{\includegraphics[width=1\linewidth]{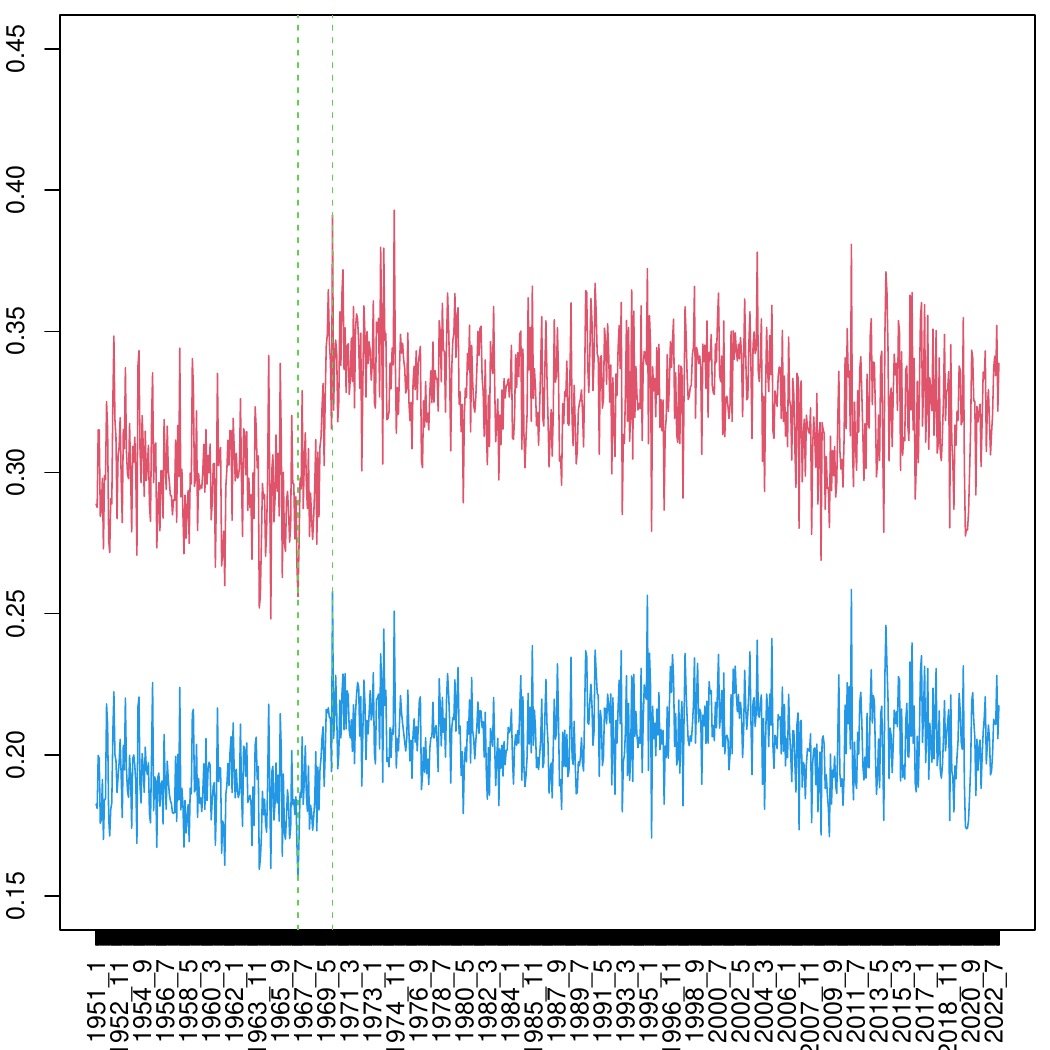}
      \label{fig:DTR_SVD12s_mnly_SB_W1_W2.pdf}}
  \end{subfigure}
  \begin{subfigure}{.66\textwidth}
  \centering
    \includegraphics[width=1\linewidth]{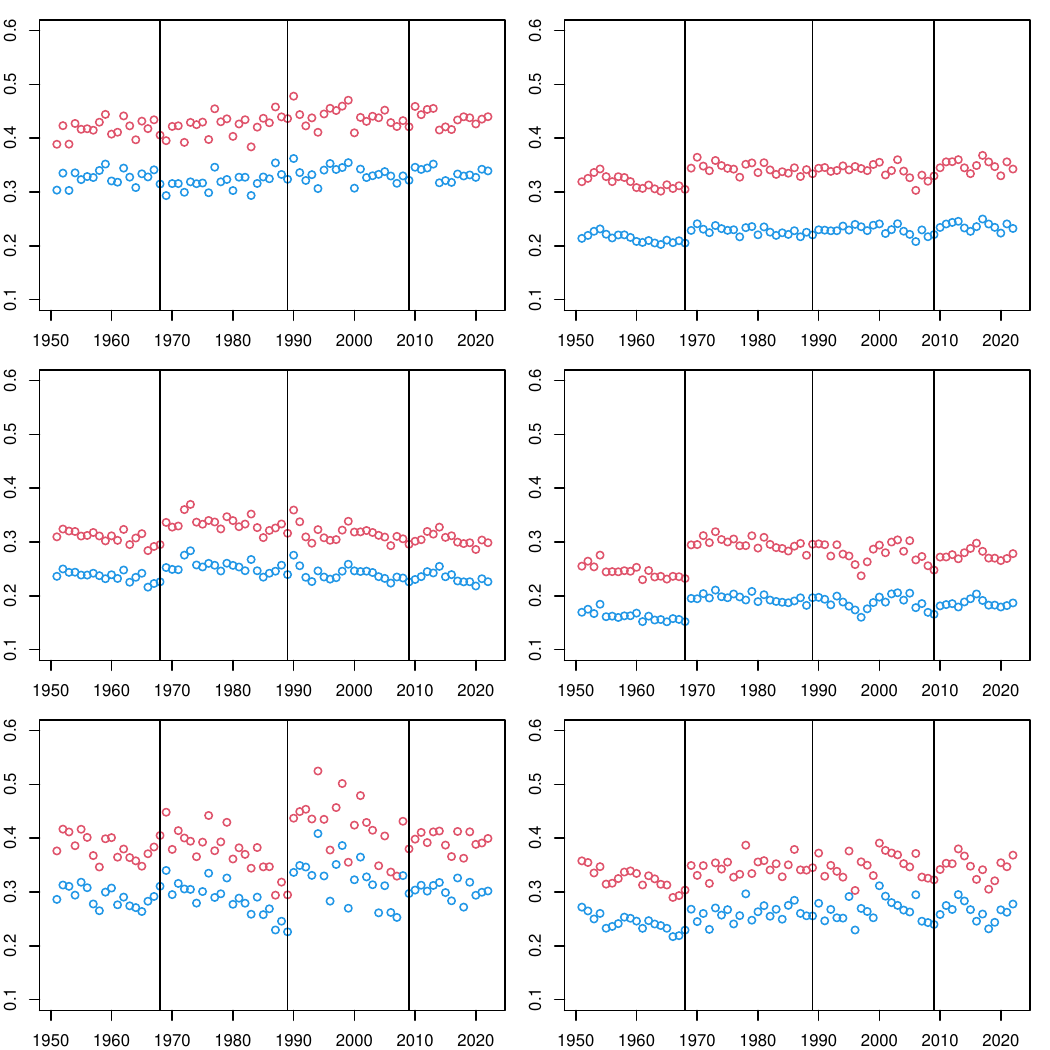}
    \caption{Yearly climatic region level data}
    \label{fig:SVD_yrly_clmtc_reg_SVD10s_DTR}
  \end{subfigure}
\caption{Spatial Bergsma measure with two spatial weight matrices for $S$:
\textit{red},  lag-1 adjacency; \textit{blue}, exponential distance decay.}
\label{fig:SB_stat}
\end{figure}

Figure \ref{fig:SB_stat} presents the $S_B$ statistics at the all India level for each year based on $S$ along with the same for each of the six climatic regions. It illustrates that both degree of association and variation in association, are climatic region specific.
The $S_B$ statistic shows without any doubt that there has been at least one, and possibly more,  major change(s) in spatial association pattern over the Indian region. Such changes in association has had varying degrees of effect in different climatic regions.

\subsection{El Ni\~no Southern Oscillation and spatial association} Figures  \ref{fig:Spiral_ord_clmtc_reg_corr_mat_with_MPdns} and \ref{fig:SB_stat} have already established that climatic regions have different impacts on the centrality and the dispersion of the spatial association measure $S_B$ for India. It is well known that El Ni\~no-Southern Oscillation (ENSO) is a key global weather phenomenon, and affects the climate of much of the tropics and subtropics (see https://www.noaa.gov/ for more information). 

\begin{figure}
    \centering
    \includegraphics[width=0.5\linewidth, height=2in]{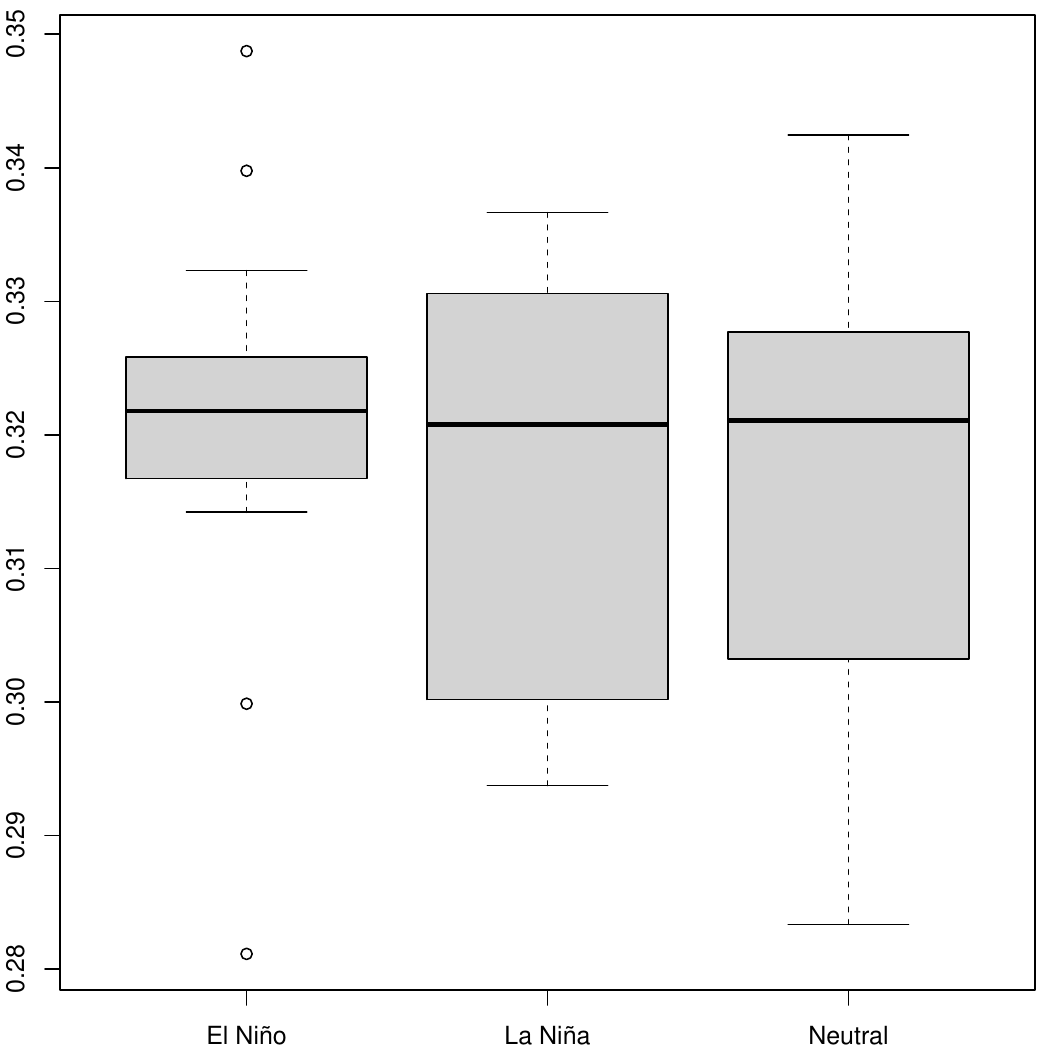}
    \caption{Boxplot of spatial Bergsma statistics, based on individual year's $S$ and categorized according to ENSO information.}
    \label{fig:ENSO_SB_lag1_SVD10s_DTR}
\end{figure}

We augmented the yearly $S_B$ statistics values obtained in the earlier section with the ENSO data. Figure \ref{fig:ENSO_SB_lag1_SVD10s_DTR} shows the impact of ENSO on the distributional behaviour of $S_B$. We can see that the variations in $S_B$ are much less in the El Ni\~no phase, and it can be argued that this is because the influence of the phase is so strong that it takes over and dictates the variation in DTR.

\section{Conclusions}\label{sec:conclusions} Data  explorations indicate that in order to bring out meaningful spatial association patterns, it is imperative to remove association due to predominant temporal patterns. The trimmed DTR matrux $S$, trimmed-off by the top 12 singular-values, and the corresponding singular-vectors, was found suitable as proxy for “time-detrended” series, with significant spatial signals still contained in it. Unsurprisingly, there is a strong association in the immediate vicinity for each of the grids, with patterns slightly differing in north-south and east-west directions. However such association should not be taken as indicative of causation.

Overall, the ESD based analyses support the presence of a climatic region-specific spatial association pattern. Climatic regions have had different impacts on the centrality and the dispersion of a spatial association measure. We can see that associations have also been affected by global climatic phenomenon like ENSO, in particular by the El Ni\~no phase. Influence of global factors have been reported in the literature earlier also \cite{vinnarasi2019}.

The yearly ESD based analyses clearly indicate a drastic change in the spatial association pattern in the late 1960s. The $S_B$ statistic confirms without any doubt that there has been at least one, and possibly more, major change(s) in the spatial association pattern in the Indian region. 

While most of the methodological techniques used here exist in the literature, a novel synergistic manner of employing these techniques to data can (and did) bring out new features. In the current situation it brought out yet unknown spatial association patterns and temporal breakage in it. Such a pipeline of analysis can be deployed to other investigative studies on spatio-temporal data as well.

We have obtained the null distribution of generalised singular values empirically,  and have used it profitably to assess 
the eigenvalues that have significant signals. This is due to the fact that distributional results on the generalised singular values for random data matrices appear to be absent in the literature. It is worth pursuing this from the theoretical perspective.

We are in broad agreement with the findings of other researchers that DTR showed an increase during early seventies till the turn of the millennium (see for instance \cite{mall2021}). This is clear even while accounting for spatial variation and using a more robust measure (than mean) against outliers.

In future the spatial association/dependence pattern that emerged in our analysis could help build composite spatio-temporal models, e.g. by using separable Gaussian space-time processes.

\bibliographystyle{abbrvnat}

\newpage

\setcounter{figure}{0}
\renewcommand{\figurename}{Figure}
\renewcommand{\thefigure}{S\arabic{figure}}

\setcounter{section}{0}
\renewcommand{\thesection}{S\arabic{section}}

\section{Supplementary information}\label{sec:suppl_plot}
We probe the DTR values further, using stratifications via climatic zones and seasons. The six climatic zones will be numbered as: (1) Tropical monsoon, (2) Tropical savannah, wet and dry (3) Arid, Steppe, hot, (4) Humid subtropical, (5) Montane climate, and (6) Hot deserts, Arid.

\subsection{Exploration of average DTR values by climatic zones and seasons} As mentioned earlier, we could use stratification by climatic zones, and by seasons within each climatic zone. 
\begin{figure}[ht!]   
     \centering
   \includegraphics[scale=0.6]{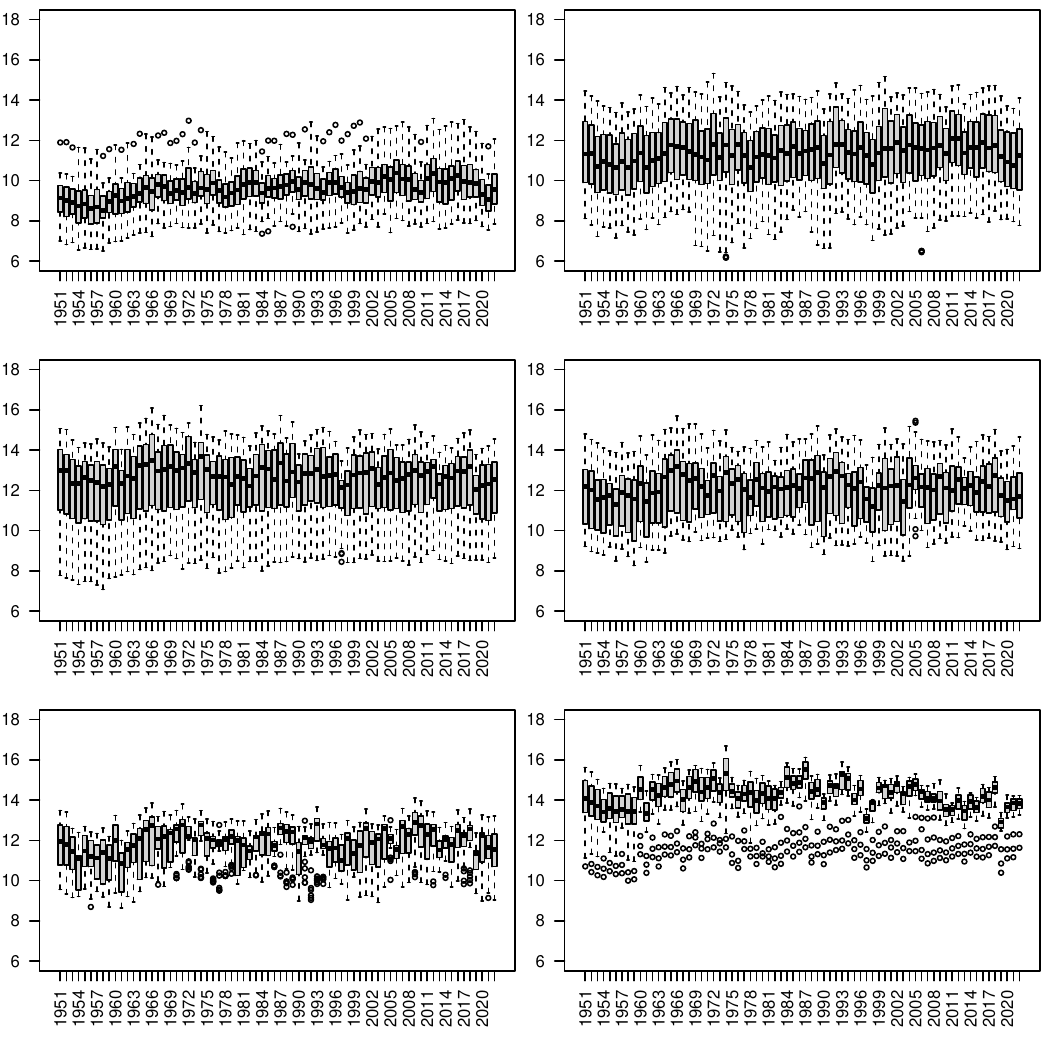}\ \ 
   \caption{Average DTR values for 72 years at the 362 locations in the six climatic zones, the six zones (1-6) presented in the order left to right and top to bottom panel.}
    \label{fig:362grid_climatic_average}
\end{figure}

\noindent \textbf{Figure \ref{fig:362grid_climatic_average}. Yearly average DTR, in each climatic zone}: As mentioned earlier, India has six climatic zones. DTR behaviour across these zones are quite different. Nevertheless, the change in pattern somewhere in the 1960’s observed earlier in the aggregated yearly data is also visible in each of these zones. There appears to be a systematic oscillation in the average DTR values for the desert climatic region. It is quite pronounced when checked at the four seasons level (approximately every 5-6 years).
\begin{figure}[ht!]
     \centering
   \includegraphics[scale=0.8]{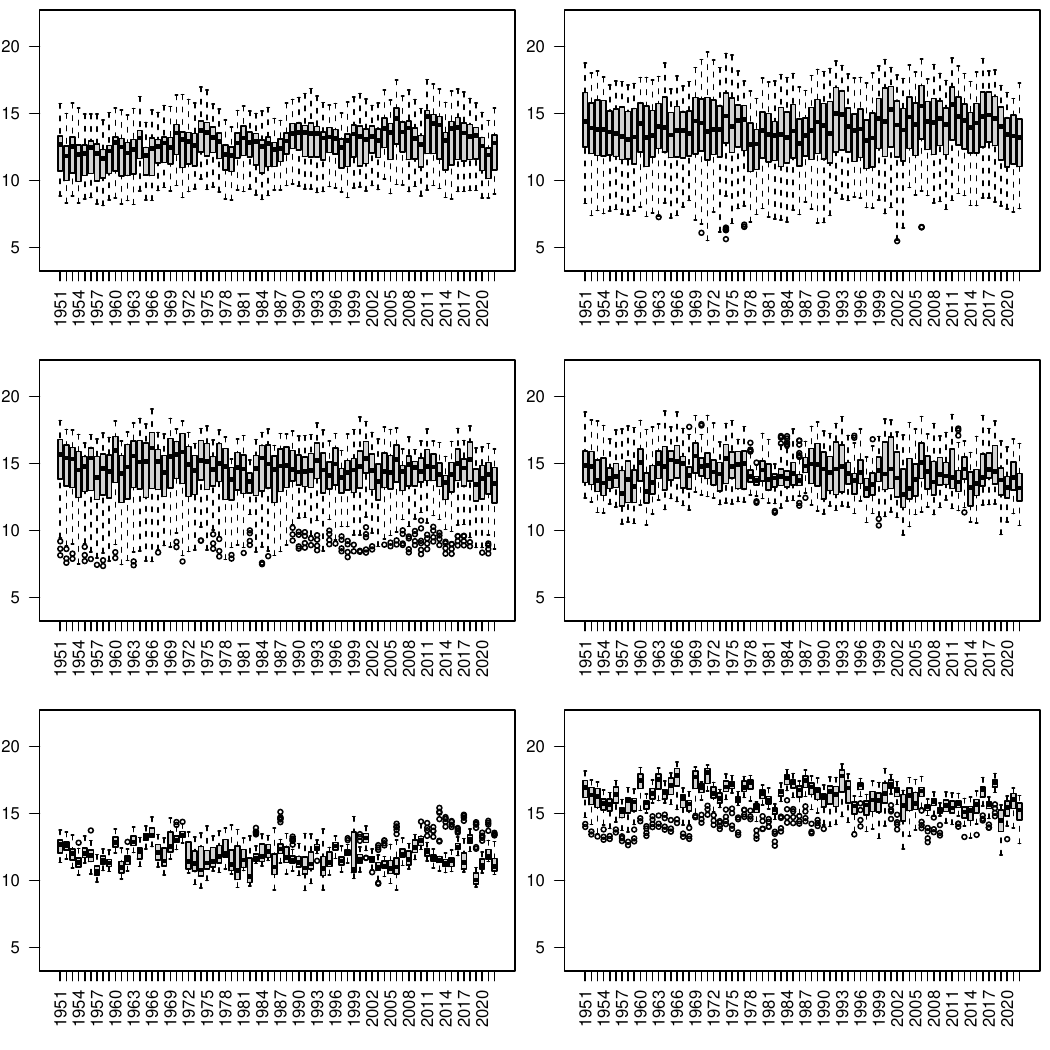}\ \    
   \caption{Average DTR values for Dec.~to Feb.~for 72 years, at the 362 locations in the six climatic zones (1)--(6) in the order left to right and top to bottom panel.}
    \label{fig:362grid_seasonal_climatic_average}
\end{figure}
\vskip5pt

\noindent \textbf{Figure \ref{fig:362grid_seasonal_climatic_average}. Yearly average DTR, in each climatic zone and season}: We can also plotting the DTR using both the seasons and the zones. This is done for the winter season (December-January-February) in this figure. Sudden change in location of the  DTR distribution is seen even at a climatic region level for any given season.

\subsection{Empirical  singular-value (SV) distribution} 
\begin{figure}[ht!]
     \centering
   \includegraphics[width=0.5\linewidth, height=2in]{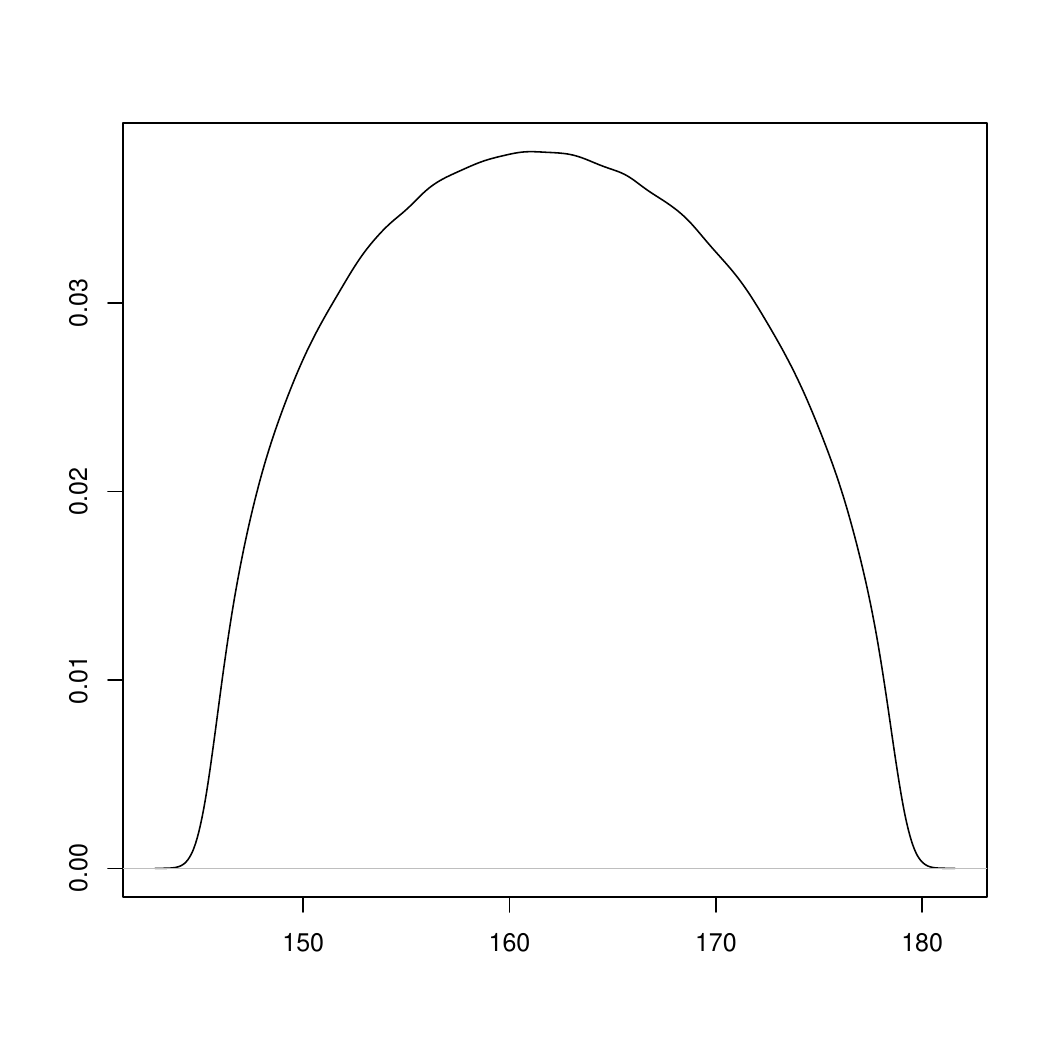}\ \    
   \caption{Empirically estimated null distribution of singular-values.}
    \label{fig:DTR_normal_null_SVs.pdf}
\end{figure}

The significant top singular-values of 
$D$ could be identified if the null distribution of singular-values for a random matrix of the same size were available. The limit distributions of such high-dimensional matrices as the dimensions grow, are known and are universal, in that they do not depend on the underlying distribution of the standardized i.i.d. entries and could be used as the null distribution. Instead, we obtained the empirical distribution of the singular-values based on 1000 simulations of matrices of order $26298\times 280$ with N(0,1) entries. Figure \ref{fig:DTR_normal_null_SVs.pdf} provides this empirical distribution. Based on the scaled DTR data, at 5\% (two-sided) significance levels, there were only 3 singular-values out of 280, that were not found significant. 

\subsection{Further findings from SVD of \texorpdfstring{$D$}{xxx}} 
\begin{figure}
    \centering
    \includegraphics[width=0.8\linewidth]{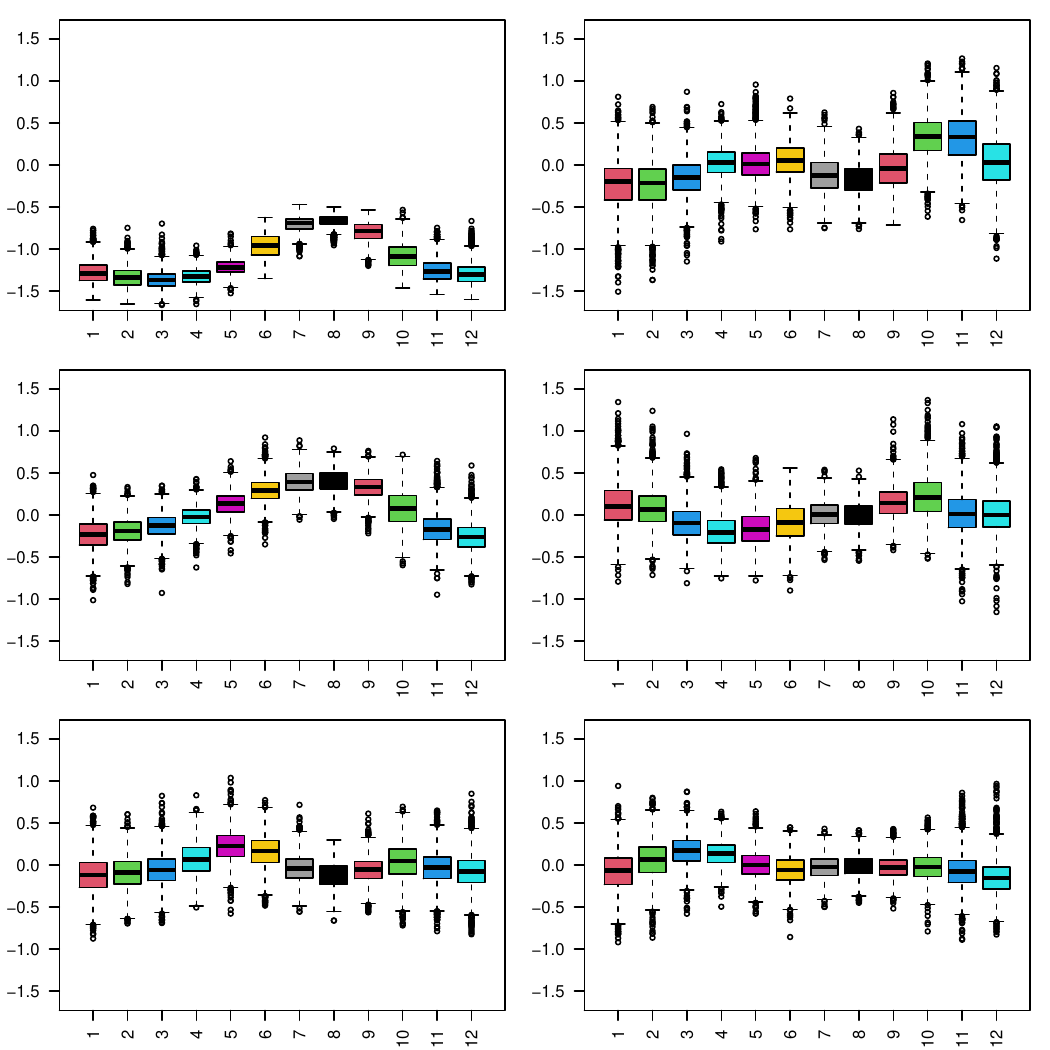}
    \caption{Components of top left singular-vectors of $D$, 
    at monthly level.}
    \label{fig:Left_SVs_and_months}
\end{figure}

\begin{figure}
    \centering
    \includegraphics[width=1\linewidth]{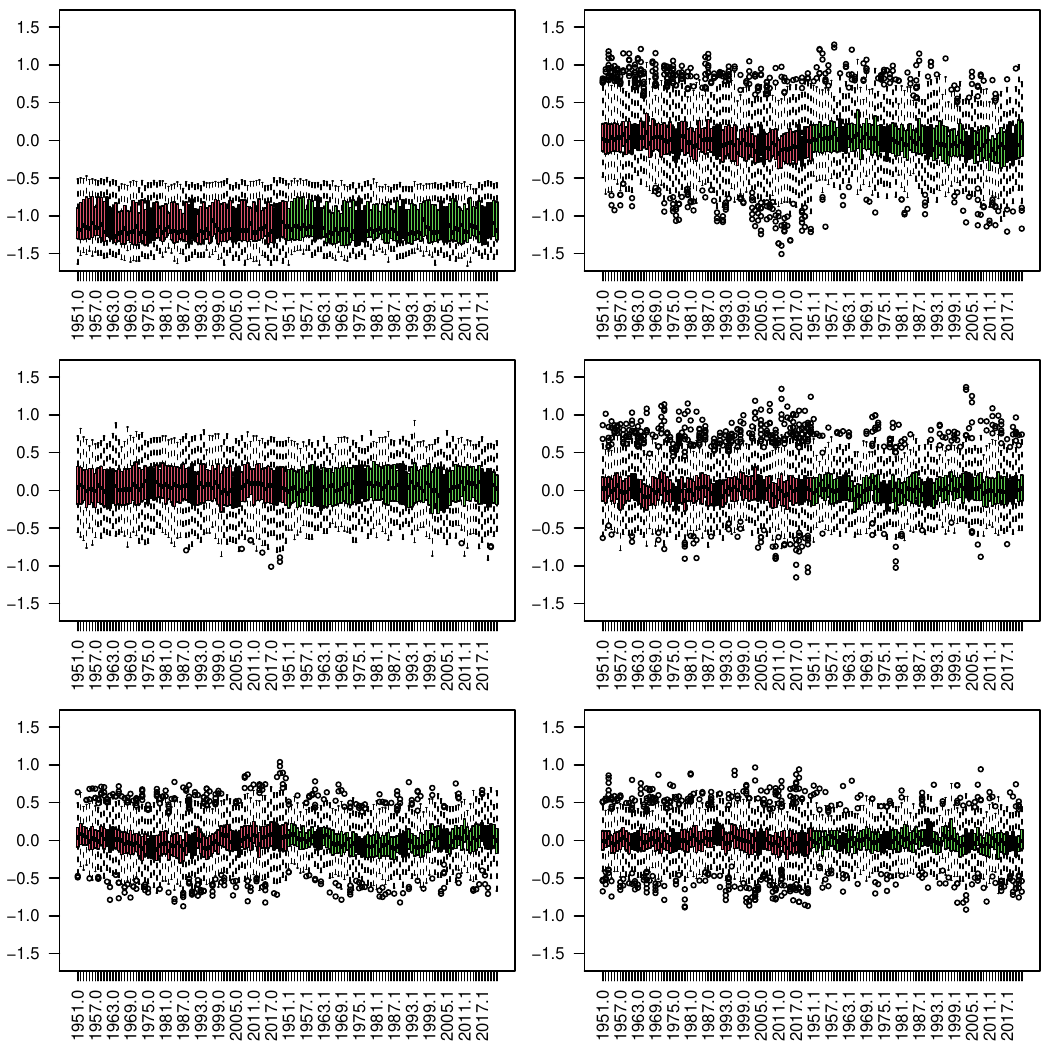}
    \caption{Components of top left singular-vectors of $D$, at weekday/weekend level for each year.}
    \label{fig:Left_SVs_and_weekends_yearly}
\end{figure}

Figures \ref{fig:Left_SVs_and_months} and \ref{fig:Left_SVs_and_weekends_yearly} present some aspects of the top 6 left singular-vectors of $D$. Unsurprisingly they have association with the seasonal patterns as captured at the month level.
It has been reported in the literature (for example, \cite{piers2003}) that weekdays and weekends can have different impacts on DTR. Figure \ref{fig:Left_SVs_and_weekends_yearly} also suggests that there could be an association between these variates.

\subsection{Eigenvalues of \texorpdfstring{$R^{D}$}{xxx}, \texorpdfstring{$R^{T}$}{xxx} and \texorpdfstring{$R^{S}$}{xxx}}
\begin{figure}
\centering
\begin{subfigure}{.3\textwidth}
  \centering
    \includegraphics[width=1\linewidth]{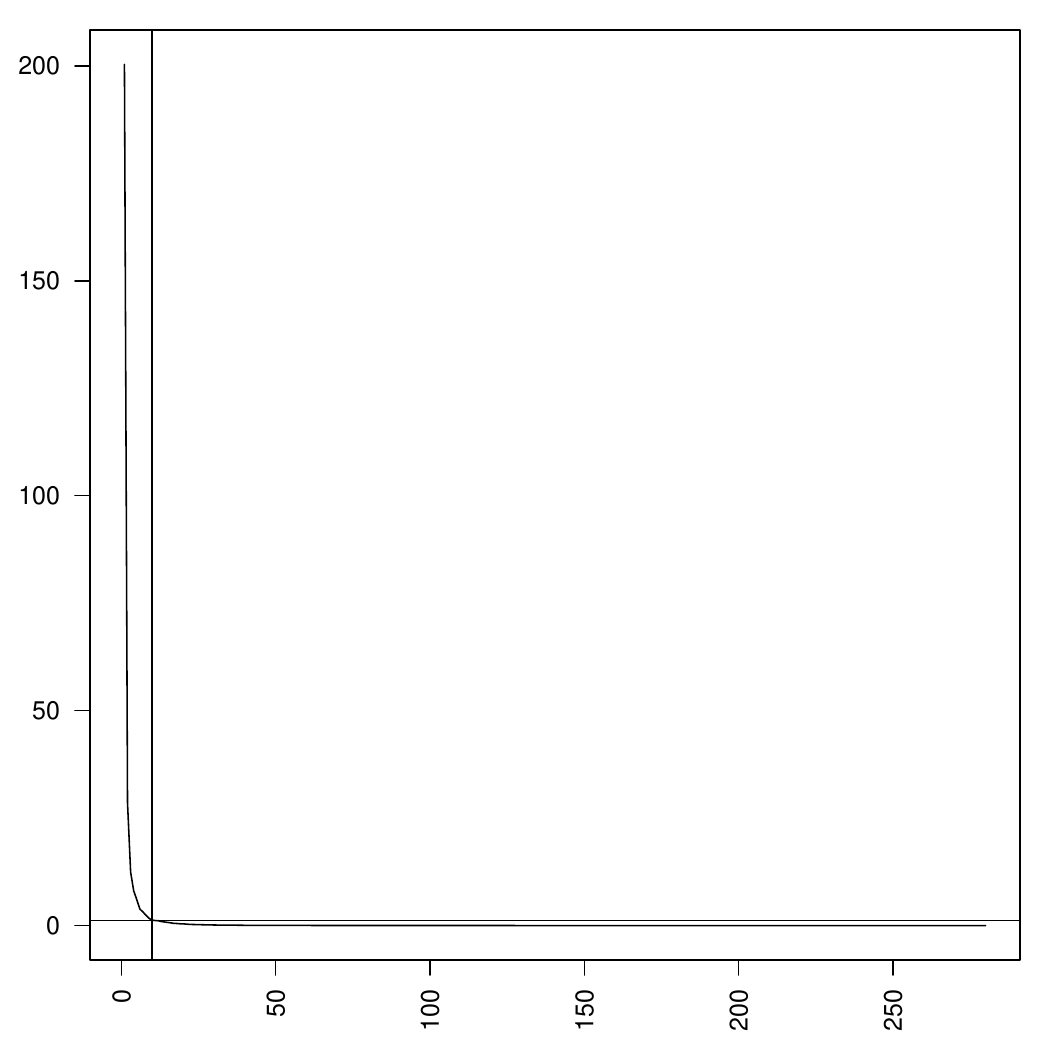}
    \caption{DTR data $D$}
  \label{fig:IMD_DTR_org_cor_EVs.pdf}
\end{subfigure}
\begin{subfigure}{.3\textwidth}
  \centering
    \includegraphics[width=1\linewidth]{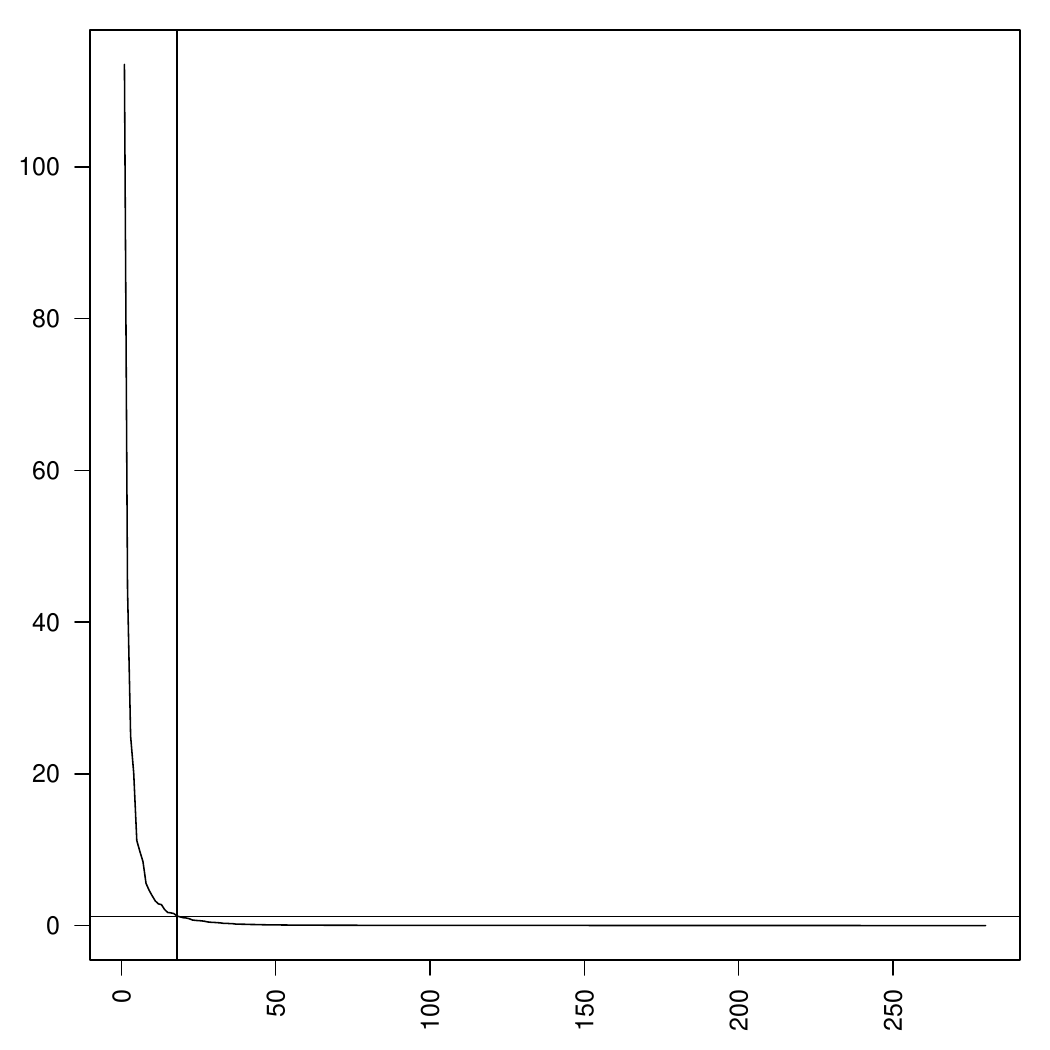}
  \caption{Time detrended data $T$.}
  \label{fig:IMD_DTR_TSdcmp_res_cor_EVs.pdf}
\end{subfigure}
\begin{subfigure}{.3\textwidth}
  \centering
    \includegraphics[width=1\linewidth]{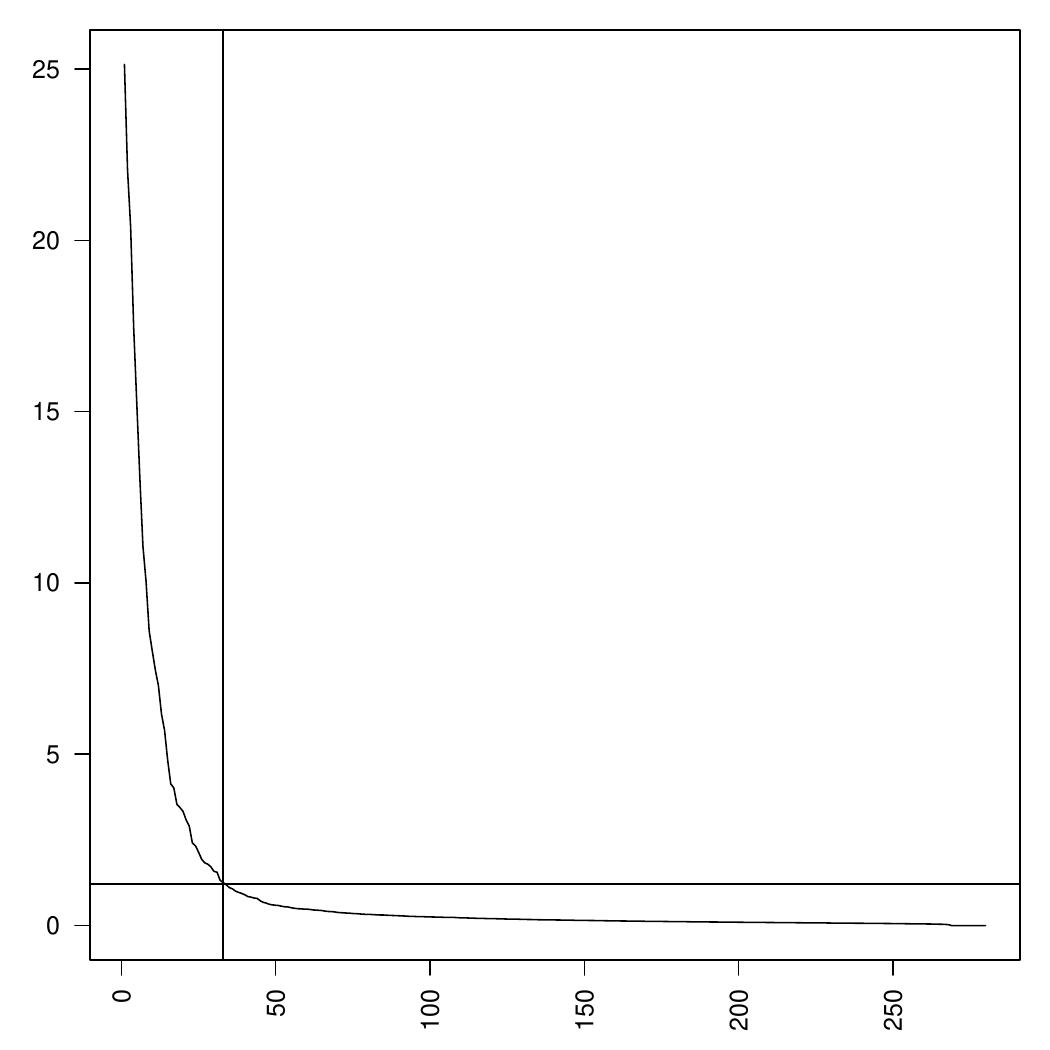}
  \caption{Trimmed data $S$.}
  \label{fig:IMD_DTR_SVD12s_res_cor_EVs.pdf}
\end{subfigure}
\caption{Eigenvalues of correlation matrices $R^{D}$, $R^{T}$ and $R^{S}$.}
\label{fig:ESDs}
\end{figure}

The spatial association of DTR over the Indian subcontinent could be studied via the ESDs from the correlation matrices based on $D$, $T$ or $S$. Figure \ref{fig:ESDs}, clearly shows presence of a handful of dominant eigenvalues for the first two, and a more reasonable decay of eigenvalues for the third. 

\subsection{Generalised singular-value decomposition (GSVD)}\label{sec:GSVD}
GSVD is a simultaneous linear transformation of two matrices $D_1$ and $D_2$ of orders $N_1\times M$ and $N_2\times M$. In implementations it is assumed that the number of rows should be more than the number of columns, so that $\min (N_1, N_2) \geq M$.

Before we give a description of GSVD let us indicate what $D_1$ and $D_2$ shall be in our context. Let the matrix $D_1$ be the $N_1 \times M$ sub-matrix of $D$ whose $n$-th row vector holds the DTR values for the $n$-th time point across all the $M=280$ grids. The $m$-th column of $D_1$ holds the DTR values of the $m$-th grid across different time points. Let $D_2$ be the $N_2\times M$ corresponding matrix for the \textit{subsequent year}. This one-to-one correspondence between the two sets of conditions is at the foundation of the GSVD comparative analysis of the two data sets.

Here we use GSVD to obtain a \textit{simultaneous linear transformation} of the two matrices $D_1$ and $D_2$ to reduced $M\times M$ space, with shared row-space basis vectors as follows: 
$$D_1 = U_1 \Delta_1 [0 \  P]V\ \ \text{and}\ \ D_2 = U_2 \Delta_2 [0  \  P]V$$ 
\noindent where $U_1$, $U_2$ and $V$ are orthogonal matrices of dimensions $N_1 \times N_1$, $N_2 \times N_2$ and $M \times M$ respectively. 

Let $r$ be the rank of the matrix $ \left[  D_1^{\top} \ D_2^{\top}\right]$, with $r \leq M$. Then $P$ is an $r \times r$ upper triangular non-singular matrix, and $[0 \ P]$ is an $r \times M$ matrix. In the case of DTR data, there is no additional information available to assume that the matrices may be rank deficient and thus it is expected that $r=M$. 

The matrices $\Delta_1$, $\Delta_2$ are non-negative quasi diagonal matrices of orders $N_1 \times r$ and $N_2 \times r$
respectively, and further,  
$$\Delta_1^\top \Delta_1 + \Delta_2^\top \Delta_2 = I,$$
\noindent where $I$ is the $r\times r$ identity matrix.

Write 
$$\Delta_1^\top \Delta_1 = {\rm diag} ( \alpha_1^2 , \ldots , \alpha_r^2 )\ \ \text{and}\ \ \Delta_2^\top \Delta_2 = {\rm diag} ( \beta_1^2 , \ldots , \beta_r^2 ),$$ where $0\leq \alpha_i, \beta_i\leq 1$ for all $i$. The ratios $\alpha_1 / \beta_1 , \ldots , \alpha_r / \beta_r$ are called the \textit{generalized singular values of the pair $(D_1, D_2)$}. If $\beta_i = 0$, then the generalized singular value $\alpha_i / \beta_i$ is infinite.

The GSVD is unique up to phase factors of $\pm1$ of each triplet of corresponding column and row basis vectors,  except in degenerate subspaces defined by subsets of pairs of generalized singular values of equal ratios. Note that the basis vectors for row spaces of the two matrices are shared (i.e. held common between the two matrices). Thus these generalised singular values capture useful essence of the two matrices relative behaviour with regard to the column spaces.

The \textit{anti-symmetric angular distances} between $D_1$ and $D_2$ are,
$$ \theta_i = \arctan(\alpha_{i}/\beta_{i}) - \pi /4, 1\leq i \leq r.$$
They indicate the relative significance of the $i$-th basis vectors from the column spaces of $D_1$ and $D_2$, and are arranged in decreasing order of significance in $D_1$ relative to $D_2$.

\subsection{Empirical generalised singular-value (GSV) distribution}\label{sec:null-GSVD}

We were unable to locate any theoretical results on distributional behaviour of the generalised singular values in literature. We propose to use Empirical GSV distribution instead. We implemented two alternate methods to simulate the null situation. In the first method, we simulate data of size similar to our current analysis setup and derive GSVs. We repeat this process 1000 times and use the collected GSVs to form an empirical distribution. In the second method, we use (random ??) permutations on the  actual data and obtain the GSVs in each case, repeating the process 1000 times, to obtain a second empirical distribution. There were no discernible differences in these two estimates of the empirical distribution of GSVs. 
\begin{figure}[ht!]
\centering
\includegraphics[width=0.6\linewidth]{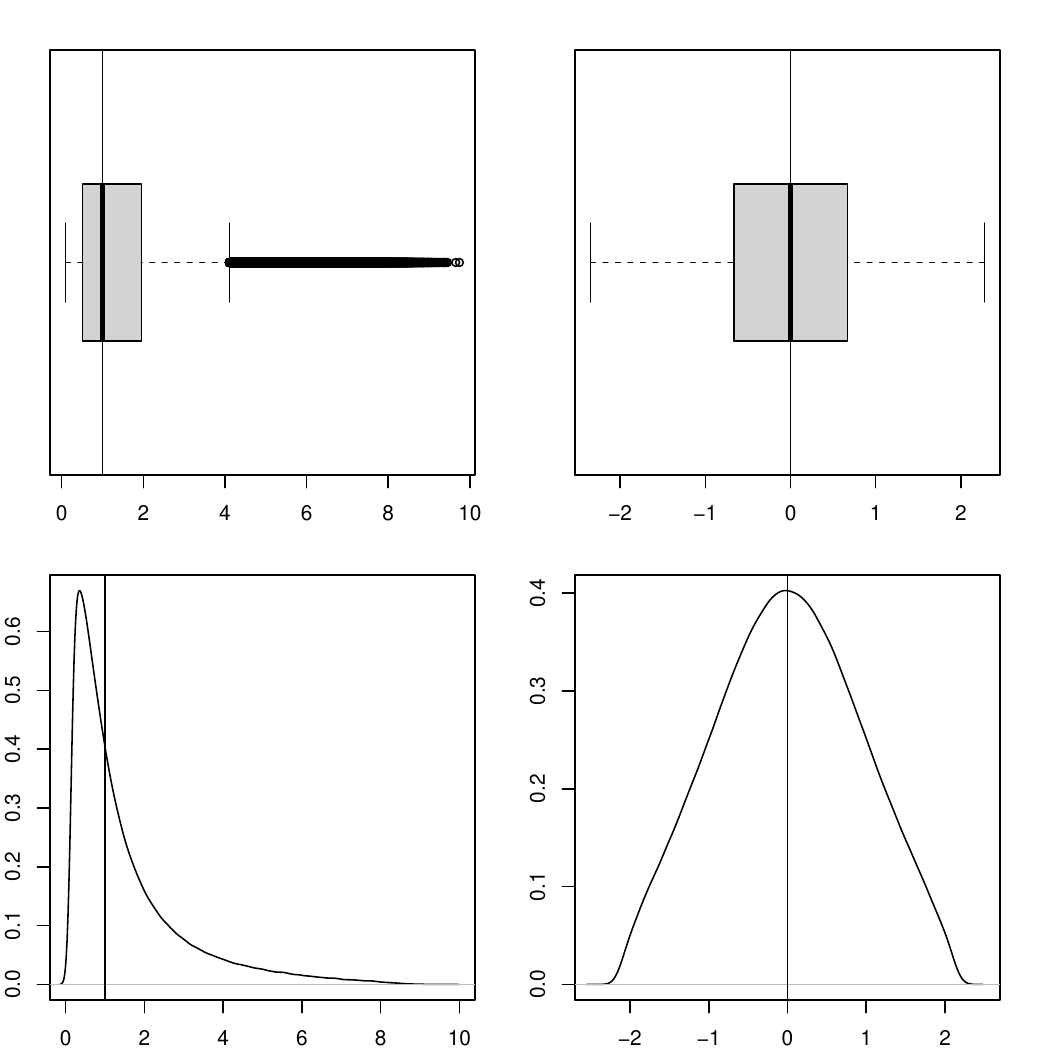}
   \caption{Simulated null distribution for  $366 \times 280$ and $365 \times 280$ matrices with independent $N(0,1)$ elements, based on 1000 simulations: \textit{left panel}--GSV; \textit{right panel}--log GSV); \textit{top panel}--boxplots;  \textit{bottom panel}--estimated overall density.}
\label{fig:nullsims}
\end{figure}
\newpage

\subsection{GSVD on DTR data and spatial information}\label{sec:sp-GSVD}
To assess retention of spatial information in $S$ compared to $D$, we could consider the sub-matrices pertaining to each year's data from $D^{\top}$ and $S^{\top}$. In that case the number of rows will be  280 from the 280 spatial locations/grids and number of columns would be 365 or 366. 

\begin{figure}
\centering
\begin{subfigure}{.45\textwidth}
  \centering
    \includegraphics[width=1\linewidth]{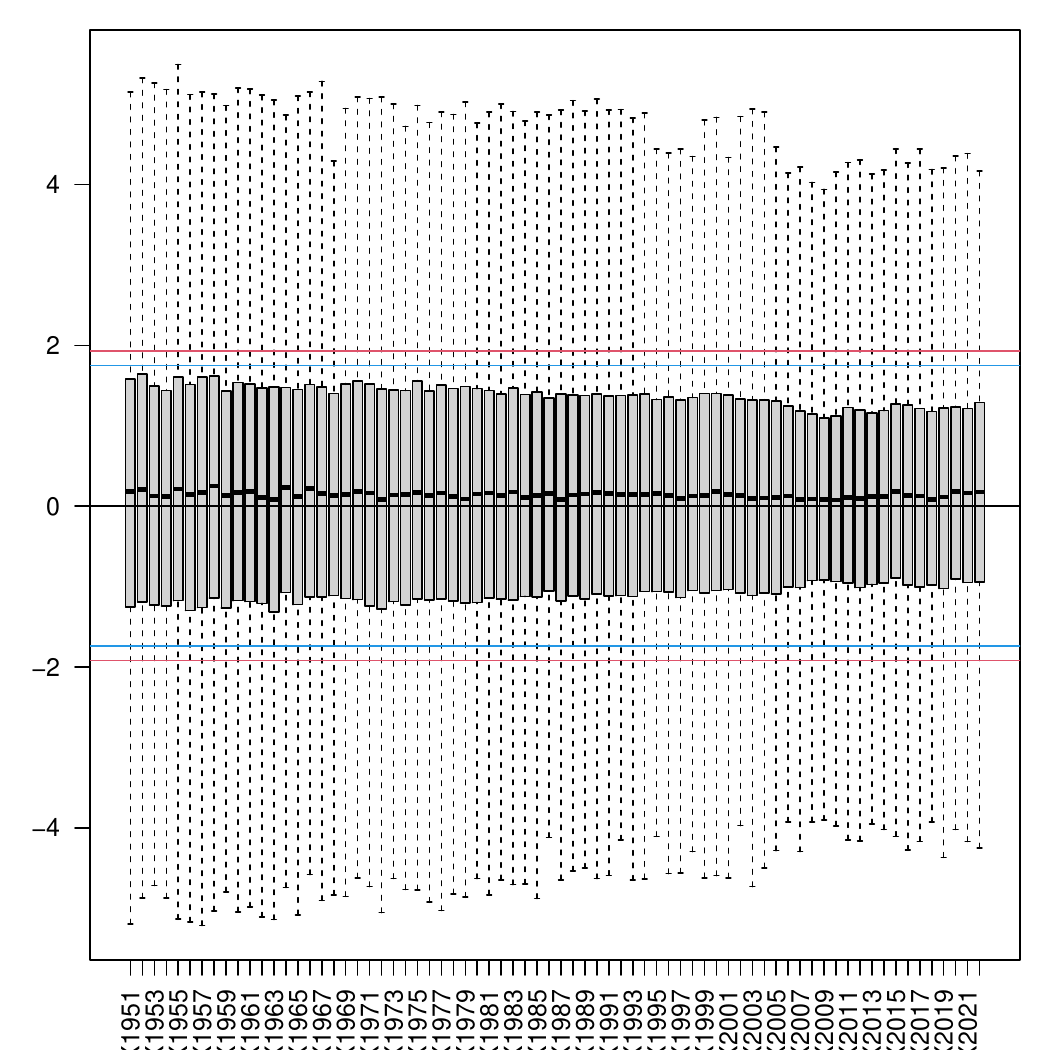}
    \caption{Original DTR data $D$.}
  \label{fig:GSVD_6mnth_org_DTR}
\end{subfigure}
\begin{subfigure}{.45\textwidth}
  \centering
    \includegraphics[width=1\linewidth]{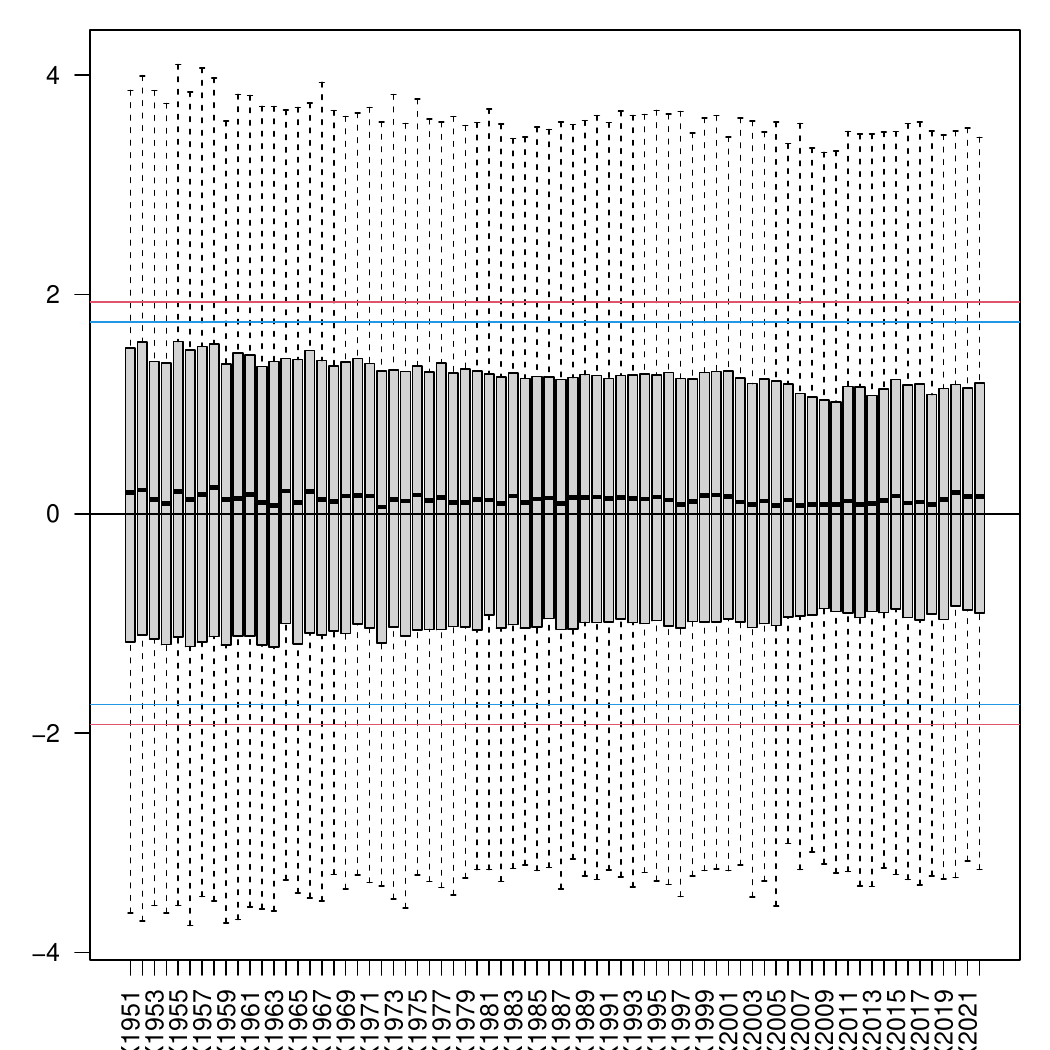}
  \caption{Trimmed DTR data $S$.}
  \label{fig:GSVD_6mnth_SVD10s_DTR}
\end{subfigure}
\caption{Empirical distributions of the (log) generalised singular-values based on transposed half-yearly data (i.e. in grid x day organisation), with original DTR data in left panel and SVD-based-trimmed DTR data in the right panel. The (empirical) critical values for the generalised SVs are marked by horizontal lines.}
\label{fig:GSVDs_6mnth}
\end{figure}

However, a technical requirement in the computation of GSVD is that the number of columns be less than the number of rows for the pair of matrices being decomposed. Thus we can not simply use the transposed matrices $D_1^\top$ and $D_2^\top$ and proceed as before. So we consider each year's data matrix $D_i$ and $S_i$ and further split it into two equal sized matrices with row sizes 182 (or 183) corresponding to the two halves of a calendar year, ignoring a singleton row for non-leap-years and consider the transpose these matrices. The results for GSVs on these ``transpose matrices'' are given in Figure \ref{fig:GSVDs_6mnth}.
The end conclusion remains as before (see Figure \ref{fig:GSVDs}), Figure \ref{fig:GSVDs_6mnth} also shows that the empirical distributions of the generalised singular-values remain same or similar for the two matrices $D$ and $S$ in the spatial direction as well.

\subsection{Analysis of alternate DTR data for India}\label{sec:alt-IN}
We employed the same analysis pipeline on an alternate data source for DTR for the Indian region from http://www.cru.uea.ac.uk/cru/data. We gathered monthly DTR data for the same 72 years and same 1 degree resolution in each of east-west and north-south direction. There were 258 grids covering India with complete data for all 72 years.
\begin{figure}[ht!]
\centering
\includegraphics[width=0.6\linewidth]{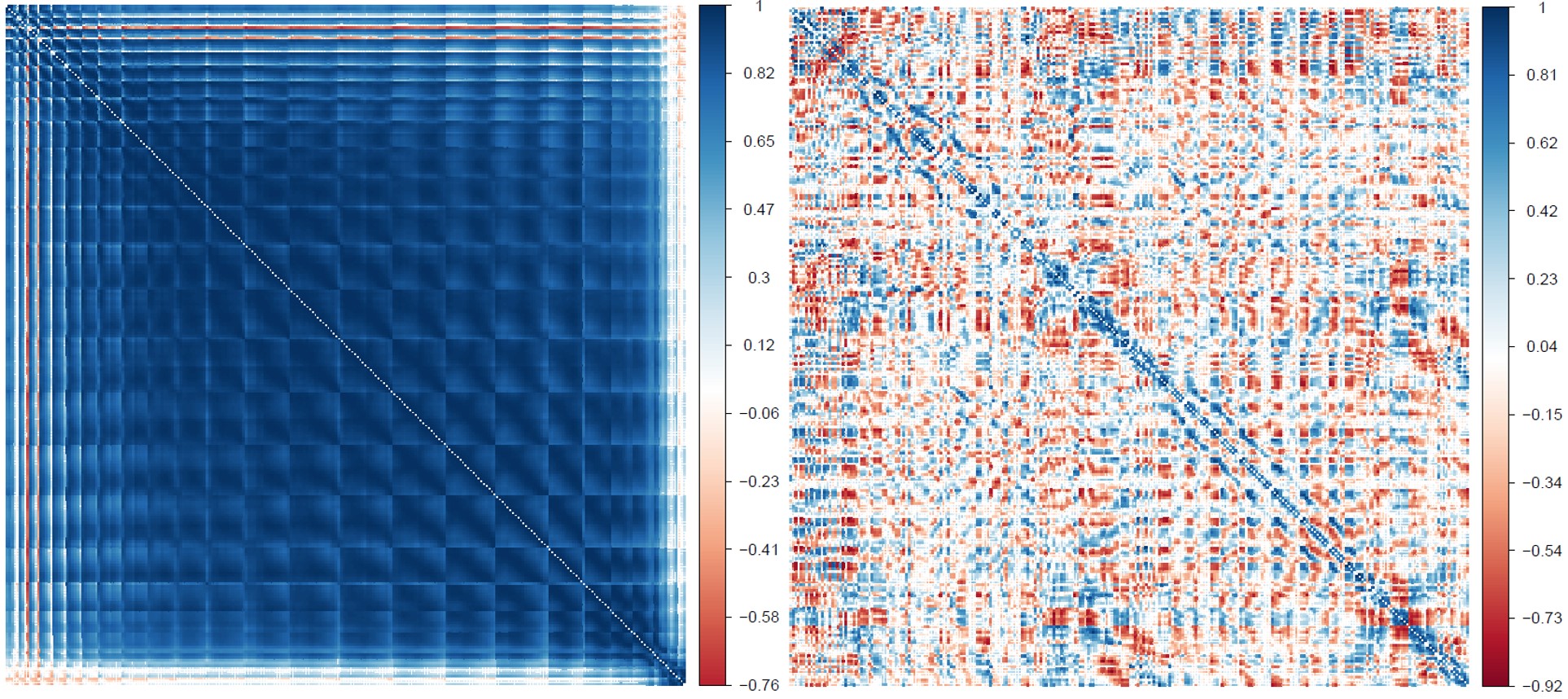}
   \caption{Correlation matrices based on monthly DTR data of India from CRU. \textit{left panel}--original DTR data; 
   \textit{right panel}--trimmed DTR with 12 singular values and grids arranged according to climatic regions).}
\label{fig:cru-cor}
\end{figure}
Figure \ref{fig:cru-cor} shows the same spatial pattern from this monthly DTR data as was identified by the daily DTR data. This provides substantial confidence in the novel analysis pipeline proposed here.

\subsection{Analysis of other geographic location: Bahia-Brazil}\label{sec:alt-BR}
Daily DTR data from Bahia of Brazil for the period January 01, 2000 to Decemebr 31, 2022 was considered. This data is not gridded, unlike the data from India considered in this manuscript. Instead, each of the 417 municipalities in Bahia contributes to an individual time series. Overall coverage is approximately about 10 degrees in both east-west and north-south direction. Thus overall this data covers a relatively smaller area with higher number of time series.
\begin{figure}[ht!]
\centering
\includegraphics[width=0.6\linewidth]{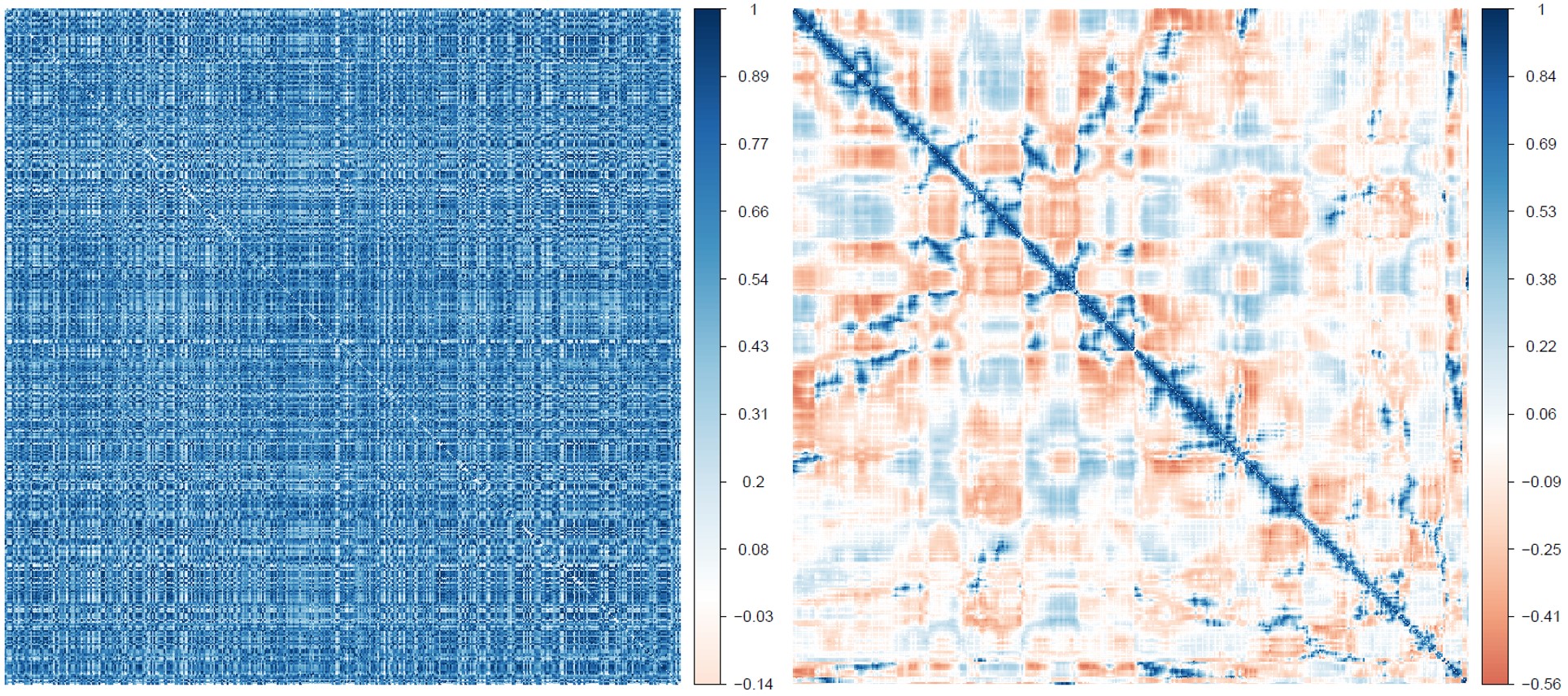}
   \caption{Correlation matrices based on daily DTR data from 417 municipalities of Bahia, Brazil. \textit{left panel}--original DTR data; \textit{right panel}--trimmed DTR with 5 singular values with municipalities arranged according to spatial proximity).}
\label{fig:bahia-cor}
\end{figure}
Figure \ref{fig:bahia-cor} elucidates novel spatial pattern for Bahia, although the underlying data is neither gridded, nor at similar spatial resolution. Also temporal coverage in this data is much shorter (only 23 years) and climatological coverage is also limited, since Bahia has mostly tropical wet and dry or savanna climate (which is one of 6 climatic regions in India).

\section{Notes on figures}\label{sec:notes_on_figures}  
We present below, at one place,  a description of all the figures given in this article. 
``DTR'' data is primarily in a matrix form $((X_{ij}))_{i \in t, j \in s}$, where $t$ is an index set for time, which could represent day/season/year, and $s$ is an index set for space, which could be the grids of size $1^\circ\times  1^\circ$ that cover India, or a climatic subset of those. The ``DTR'' data could be the original DTR data $D$, the time series decomposition based detrended DTR data $T$, or the SVD based detrended trimmed DTR data $S$. 

For original daily DTR data, $D$, $t$ spans 26298 days covering 1-Jan-1951 to 31-Dec-2022 and $s$ covers the 362 grids that cover India. We have complete data for the 72 years (1951-2022) for 280 of these grids, so often the size of $s$ is 280. However $s$ could also represent one of the six climatic zones of India.

In various figures DTR values have been summarised using functions like average over a (sub)set of time and/or space, correlation matrices, empirical spectral distributions, singular-values, singular-vectors, eigenvalues, generalised singular-values, etc. We use the following notation to describe the figures: 
\vskip3pt

\noindent $D=((d_{ij}))_{26298\times 280}$ is the data matrix. 
\vskip3pt

\noindent
$U \Sigma V^\top=D$ is the singular-value decomposition, where $\Sigma = Diag(\sigma_j, j=1, \ldots, 280)$, and the singular-values $\{\sigma_j\}$'s are arranged in decreasing order. 
\vskip3pt

\noindent $S= ((S_{ij}))_{26298\times 280}=D - \sum_{j=1}^{12} \sigma_j  u_j v_j^\top$ is the trimmed data matrix.
\vskip3pt

\noindent $T= ((S_{ij}))_{26298\times 280}$ 
is the residual matrix after a time series decomposition of each column of $D$ using additive models each consisting of a trend component (estimated by simple moving average), a seasonal component and a random/residual component.
\vskip3pt

\noindent $R^D$ is the $280\times 280$ correlation matrix based on $D$.
\vskip3pt

\noindent $R^{D}_i$ is the $280\times 280$ correlation matrix based on $D$ for the $i$-th year.
\vskip3pt

\noindent $R^{S}$ is the $280\times 280$ correlation matrix based on $S$.
\vskip3pt

\noindent $R^{S}_i$ is the $280\times 280$ correlation matrix based on $S$ for the $i$-th year.
\vskip3pt

\noindent 
$\{\lambda^{D}_j\}$ and $\{\lambda^{S}_j\}$ are the eigenvalues, in decreasing order, of the correlation matrices $R^{D}$ and $R^{S}$. 
\vskip3pt

\noindent $\hat{R}^{D}:= \sum_{j=1}^{10} \lambda^{D}_{j}  e^{D}_{j} {(e^{D}_{j})}^\top$ is the 
de-noised approximation of $R^D$.
\vskip3pt

\noindent $\hat{R}^{S} := \sum_{j=1}^{33} \lambda^{S}_{j}  e^{S}_{j} {(e^{S}_{j})}\top$ is the 
de-noised approximation of $R^S$.
\vskip3pt

\noindent $SB^{S}_{i}$: Spatial Bergsma statistic using $S$, for the $i$-th year.
\vskip3pt

\noindent $SB^{S}_{ij}$: Spatial Bergsma statistic using $S$, for the $i$-th year in the $j$-th climatic zone. 
\vspace*{0.2in}

\noindent 
\textbf{Figure \ref{fig:India_climate_map.pdf}}: India divided into 362 $1^\circ\times  1^\circ$ grids, each grid labelled by one of the six climatic region codes.
\vspace*{0.15in}

\noindent 
\textbf{Figure \ref{fig:362grid_dtr_yearly_average}}: 
Boxplots for each of the 72 years, of the yearly averages from the 362 grids.
\vspace*{0.15in}

\noindent 
\textbf{Figure \ref{fig:362grid_seasonal_average}}: Boxplots for averages over the daily data from $D$, for each given season (DJF, MAM, JJA, SON), year ($i=1, \ldots, 72$) and grid ($j =1, \ldots 362$).
\vspace*{0.15in}

\noindent 
\textbf{Figure \ref{fig:Eigenvalues_of_yrly_cor_mat_of_org_DTR}}:
For every year $i$, the left panel are the bulk eigenvalues (presented in the form of boxplots without outliers) of $R^{D}_i$, 
The right panel 
is the plot of percentiles, $0(10)100$, of these eigenvalues. 
\vspace*{0.15in}

\noindent 
\textbf{Figure \ref{fig:DTR_cor_MPdns}}.
Heat map: the upper triangle is based on the correlations from $R^D$; the lower triangle is for the corresponding entries from the matrix $\hat{R}^{D}$.

\vspace*{0.15in}

\noindent 
\textbf{Figure \ref{fig:Singular_values_of_org_DTR}}: 
Plot of the cumulative sums of the singular-values of $D$. 
\vspace*{0.15in}

\noindent 
\textbf{Figure \ref{fig:ACF_of_org_DTR_data}}: 
$280$ auto-correlation series for each column of $S$ treated as a time series.
\vspace*{0.15in}

\noindent 
\textbf{Figure \ref{fig:ESD_org_DTR_and_SVR10res}}: Box plots of the eigenvalues $\{\lambda^{D}_j\}$ and $\{\lambda^{S}_j\}$ of  $R^{D}$ and $R^{S}$.
\vspace*{0.15in}

\noindent 
\textbf{Figure \ref{fig:GSVDs}}: Consider the daily trimmed DTR matrix $S$. Let $S^{y_1}$ and $S^{y_2}$ be data on two successive years ($y_1$ and $y_2$) from  $S$. GSVD enables simultaneous decomposition of these matrices as 
$$S^{y_1} = U_1 \Delta_1 [0 \, \, P]V\ \ \text{and}\ \ S^{y_2} = U_2 \Delta_2 [0 \, \, P]V$$ where $U_1$, $U_2$ and $V$ are orthogonal matrices, $\Delta_1$, $\Delta_2$ are non-negative quasi diagonal matrices which satisfy $\Delta_1^\top \Delta_1 + \Delta_2^\top \Delta_2 = I$ and $P$ is an upper triangular non-singular matrix. Then generalised singular-values (gsv) are the ratios of the respective elements from the diagonals of $\Delta_1$ and $\Delta_2$. Typical implementation of this algorithm ensures these ratios are positive and decreasing. 
The figure gives 71 boxplots of these gsvs for each successive pairs of years. The figure also includes similar boxplots for the gsvs based on $D$.
\vspace*{0.15in}

\noindent 
\textbf{Figure \ref{fig:IMD_DTR_SVD12s_res_corr_4cities.pdf}}: 
For each of the four major cities, Delhi, Kolkata, Chennai and Mumbai, the correlations (from $R^S$), with the other cities have been plotted as heat maps.
\vspace*{0.15in}

\noindent 
\textbf{Figure \ref{fig:Max_cor_lat_long}}:  
Fix a grid $g_1$, and consider the grid $g_2$ which yields the maximum correlation (based on $R^{S}$) with $g_1$. Consider the difference in latitudes and longitudes between  $g_1$ and $g_2$. The two distribution of these differences across all grids have been plotted.
\vspace*{0.15in}

\noindent 
Figures \ref{fig:Bulk_ESD_yrly_cor_mat}--\ref{fig:Spiral_ord_clmtc_reg_corr_mat_with_MPdns} are similar to Figures \ref{fig:ESD_of_cor_mat_of_org_DTR}--\ref{fig:DTR_cor_MPdns} but based on $S$ instead of $D$.
\vspace*{0.15in}

\noindent 
\textbf{Figure \ref{fig:Spiral_ord_clmtc_reg_corr_mat_with_MPdns}}: Heat map: the upper triangle is for the correlations from $R^S$; the lower triangle is for the corresponding entries from the matrix $\hat{R}^{S}$. For both matrices, the grids are arranged first by climatic regions, and within each region in a spiral manner according to a Hilbert space filling curves.
\vspace*{0.15in}

\noindent 
\textbf{Figure \ref{fig:Bulk_ESD_yrly_cor_mat}}: For every year $i$, plot of bulk of the eigenvalues (as boxplot without outliers) of $R^{S}_i$. 
\vspace*{0.15in}

\noindent 
\textbf{Figure \ref{fig:SB_stat}}: Spatial Bergsma statistics $S_B$ with two spatial weight matrices, $W_1$ (lag-1 adjacency, in red) and $W_2$ (exponential distance decay in blue), for $S$. In the left panel, the yearly $SB^{S}_{i}$, $i=1, \ldots, 72$. In the right panel, the yearly $SB^{S}_{ij}$, for $1\leq i \leq  72, 1\leq j \leq 6$.
\vspace*{0.15in}

\noindent 
\textbf{Figure \ref{fig:ENSO_SB_lag1_SVD10s_DTR}}: Boxplots of $\{SB^{S}_{i}\}$, $i=1, \ldots, 72$ for the three ENSO categories. 
\vspace*{0.15in}

\noindent 
\textbf{Figure \ref{fig:362grid_climatic_average}}: Average DTR values for each of the 72 years at the 362 locations in the six climatic zones, the six zones (1-6) presented in the order left to right and top to bottom panel.
\vspace*{0.15in}

\noindent 
\textbf{Figure \ref{fig:362grid_seasonal_climatic_average}}: Average DTR values in the winter season (December to February) for each of the 72 years, at the 362 locations in the six climatic zones, with the six zones (1-6) presented in the order left to right and top to bottom panel.
\vspace*{0.15in}

\noindent 
\textbf{Figure \ref{fig:DTR_normal_null_SVs.pdf}}: Simulated null distribution of Singular Values based on matrices of dimensions $26298 \times 280$ with independent $N(0,1)$ elements and 1000 replications.
\vspace*{0.15in}

\noindent 
\textbf{Figure \ref{fig:Left_SVs_and_months}}: Components of the top 6 left singular-vectors of $D$, split according to months, yielding $6\times 12$ boxplots. 
\vspace*{0.15in}

\noindent 
\textbf{Figure \ref{fig:Left_SVs_and_weekends_yearly}}: Components of the top 6 left singular-vectors of $D$, split according to year, and within each year by weekday or weekend, yielding $72\times 2$ plots for each component. 
\vspace*{0.15in}

\noindent 
\textbf{Figure \ref{fig:ESDs}}: Eigenvalues of correlation matrices, based on $D$, $S$ and $T$.
\vspace*{0.15in}

\noindent 
\textbf{Figure \ref{fig:nullsims}}: Simulated null distribution of Generalised Singular Values (\textit{left panel}) and log GSV (\textit{right panel}), with boxplots from each of 1000 simulations on top panel and estimated overall density in the bottom panel. 
Based on matrices of dimensions $366 \times 280$ and $365 \times 280$ with independent $N(0,1)$ entries.
\vspace*{0.15in}

\noindent 
\textbf{Figure \ref{fig:GSVDs_6mnth}}: For each year $i$, split $S_i^\top$ into two approximately equal halves and perform GSV decomposition on the resulting two 280 x 182 (or 183) dimensional matrices. Figure shows boxplots of these GSV's from each year. Also included are similar boxplots based on $D_i^\top$.
\vspace*{0.15in}

\noindent 
\textbf{Figure \ref{fig:cru-cor}}: This figure is similar to Figures \ref{fig:DTR_cor_MPdns} and \ref{fig:Spiral_ord_clmtc_reg_corr_mat_with_MPdns} but with alternate data source and is based on monthly data instead of daily DTR data from India (and without MP de-noising). 
\vspace*{0.15in}

\noindent 
\textbf{Figure \ref{fig:bahia-cor}}: This figure is similar to Figures \ref{fig:DTR_cor_MPdns} and \ref{fig:Spiral_ord_clmtc_reg_corr_mat_with_MPdns} but with data from Bahia, Brazil (and without MP de-noising).

\end{document}